\documentclass[a4paper,amsmath,amssymb,color,nofootinbib,preprint,tightenlines]{revtex4-2}
\usepackage{graphicx}
\usepackage[dvipsnames]{xcolor}
\usepackage[mathscr]{euscript}
\usepackage{slashed}
\usepackage{tablefootnote}
\usepackage[normalem]{ulem}
\usepackage[colorlinks=true,pdfstartview=FitV,citecolor=blue,linkcolor=purple,urlcolor=Sepia]{hyperref}

\begin{document}
\baselineskip=17pt \parskip=3pt

\title{\boldmath$B\to K\rm{+}\,invisible$, dark matter, and \\
$CP$ violation in hyperon decays}

\author{Xiao-Gang He$,^{1,}$\footnote{\email{hexg@sjtu.edu.cn}} Xiao-Dong Ma$,^{2,3,}$\footnote{\email{maxid@scnu.edu.cn}} Jusak Tandean$,^{1,}$\footnote{\email{jtandean@yahoo.com}} and German Valencia$^{4,}$\footnote{\email{german.valencia@monash.edu}} \vspace{1ex} \\ \it
$^1$State Key Laboratory of Dark Matter Physics, Tsung-Dao Lee Institute and School of Physics and Astronomy,
Shanghai Jiao Tong University, Shanghai 201210, China \vspace{1ex} \\
$^2$State Key Laboratory of Nuclear Physics and Technology, Institute of Quantum Matter, South China Normal University, Guangzhou 510006, China \vspace{1ex} \\
$^3$Guangdong Basic Research Center of Excellence for Structure and Fundamental Interactions of Matter, Guangdong Provincial Key Laboratory of Nuclear Science, Guangzhou 510006, China\vspace{1ex} \\
$^4$School of Physics and Astronomy, Monash University, Wellington Road, Clayton, Victoria 3800, Australia \vspace{2em} \\
\rm Abstract \vspace{8pt} \\
\begin{minipage}{0.99\textwidth} \baselineskip=17pt \parindent=3ex \small
Recently the Belle II Collaboration has reported a measurement of the $B^+\to K^+\nu\bar\nu$ rate that is higher than the standard-model expectation.
Since the emitted neutrinos are unobserved, the excess could be due to the $B^+$ decaying into a $K^+$ and a dark-matter pair.
We entertain this possibility in a two-Higgs-doublet model supplemented with a real singlet scalar boson acting as the dark matter.
This model also accommodates strangeness-changing interactions providing new sources of $CP$ violation which can affect hyperon and kaon  nonleptonic transitions.
We find that the resulting $CP$ violation in the hyperon sector can be significant, reaching the current empirical bounds, after taking into account constraints from kaon mixing and decay and from dark-matter relic-density data and direct searches including the Migdal effect.
We demonstrate that the hyperon and kaon processes are complementary probes of this new-physics scenario.
Its prediction for sizable hyperon $CP$ violation is potentially testable in ongoing experiments, such as BESIII, Belle II, and LHCb, and in next-generation ones like PANDA and at the Super Tau Charm Facility.
\end{minipage}}

\maketitle

\newpage

\section{Introduction\label{intro}}

Rare processes are important probes of new physics (NP) beyond the standard model (SM). Although the rareness makes their detection difficult, they have an advantage in that NP may increase their rates substantially above the SM backgrounds and sufficiently to be discoverable when experiments with high sensitivity become available.

Among the most interesting rare decays are those of hadrons with invisible particles in the final state.
One example is \,$B^+\to K^+\rm{+}invisible$,\, which has garnered much attention lately.
In the SM the invisibles are an undetected neutrino-antineutrino pair $(\nu\bar\nu)$, and theoretical calculations of its branching fraction are relatively precise compared to decays with charged leptons or more hadrons in the final state, resulting in \,${\cal B}(B^+ \to K^+ \nu \bar \nu)_{\rm SM} = (4.43\pm 0.31)\times 10^{-6}$~\cite{Becirevic:2023aov,He:2023bnk},
where the tau-lepton-mediated tree-level contribution has been removed.
In 2023 the Belle II experiment~\cite{Belle-II:2023esi}, after searching for this mode,  found \,${\cal B}(B^+ \to K^+ \nu \bar \nu) = (2.3\pm 0.7)\times 10^{-5}$, exceeding the SM prediction by 2.7$\sigma$.
Combining this with the earlier findings by BaBar \cite{BaBar:2010oqg,BaBar:2013npw}, Belle \cite{Belle:2013tnz,Belle:2017oht}, and Belle~II~\cite{Belle-II:2021rof} yields the weighted average \,${\cal B}(B^+ \to K^+ \nu \bar \nu)_{\rm exp} = (1.3 \pm 0.4)\times 10^{-5}$~\cite{Belle-II:2023esi}, bigger than the SM value by 2.1$\sigma$.

Needless to say, this excess still has to be confirmed with more measurements.
If it persists with increasing statistical significance, it will constitute evidence for NP beyond the SM and therefore should be investigated accordingly.

It is attractive to devise a dark matter (DM) explanation for this anomaly within a consistent model.
Since the neutrinos in the final state are not identified, unobserved particles from beyond the SM could also be emitted in \,$B^+\to K^+\rm{+}invisible$, thereby adding to its SM rate, as recently explored in different scenarios~\cite{He:2023bnk,Felkl:2023ayn,Abdughani:2023dlr,Wang:2023trd,Berezhnoy:2023rxx,Datta:2023iln,Altmannshofer:2023hkn,McKeen:2023uzo,Fridell:2023ssf,Ho:2024cwk,Gabrielli:2024wys,Hou:2024vyw,Bolton:2024egx,Buras:2024ewl,Kumar:2024ivx,Hati:2024ppg,Altmannshofer:2024kxb,Hu:2024mgf,Calibbi:2025rpx,He:2024iju}.
In the DM case, the exotic channel \,$B^+\to K^+D\bar D$\, could be open, caused by the quark transition \,$b\to sD\bar D$, where the DM pair, $D\bar D$, emerges invisibly.

One possibility entertained hereafter is that $D$ is a real scalar boson carrying no charge under the SM gauge groups, as occurred in various contexts~\cite{Bird:2006jd,Kim:2009qc,He:2010nt,Badin:2010uh,Kamenik:2011vy,He:2022ljo,He:2023bnk,He:2024iju} accommodating $b\to sDD$.
As elaborated in refs.~\cite{He:2023bnk,He:2024iju}, scenarios of this kind can account for the aforesaid Belle II excess and also furnish DM-quark couplings, $DD\textit{\texttt q}\bar{\textit{\texttt q}}'$, which induce the right amount of DM relic abundance and DM-nucleon scattering testable in DM direct-detection experiments.

In this paper, working along similar lines, we consider an ultraviolet-complete model that contains a scalar sector comprising two Higgs doublets plus the $D$ particle, generates the desired $DD\textit{\texttt q}\bar{\textit{\texttt q}}'$ couplings, and gives rise to four-quark flavor-changing neutral-current transitions at tree level.
The last feature implies that there could be new contributions to strangeness-changing processes, specifically nonleptonic hyperon and kaon ones, which were not discussed in the previous studies.
Of great interest here is whether the model can produce $CP$ violation in hyperon decays surpassing in magnitude what the SM predicts and at levels which may be observable at currently operational and planned facilities.
Although multiple searches for hyperon $CP$ violation have been performed over the years and come up empty~\cite{Belle:2022uod,LHCb:2024tnq,R608:1985fmh,Barnes:1987vc,DM2:1988ppi,Barnes:1996si,E756:2000rge,HyperCP:2004zvh,HyperCP:2006ktj,BESIII:2018cnd,BESIII:2021ypr,BESIII:2022qax,BESIII:2022lsz,BESIII:2022lsz,BESIII:2023jhj,BESIII:2023lkg,BESIII:2023drj,BESIII:2024nif,BESIII:2020fqg,BESIII:2023sgt}, there remains open a wide window into potential NP to be inspected with already running machines, particularly BESIII \cite{BESIII:2020nme,Zheng:2025tnz}, Belle II~\cite{Belle:2022uod}, and LHCb~\cite{LHCb:2024tnq}, the upcoming  PANDA experiment at FAIR~\cite{PANDA:2020zwv}, and future efforts at the proposed Super Tau Charm Facility~\cite{Achasov:2023gey}.

The rest of the paper is arranged as follows.
In section \ref{model} we describe the salient features of the NP model we employ, especially the part of its Yukawa sector responsible for the effective DM-quark and four-quark operators to be examined.
In sections \ref{b2sdd} and \ref{DMsector} we discuss how the model can, respectively, explain the \,$B^+\to K^+\rm{+}invisible$\, anomaly and reproduce the observed DM relic density without conflicting with DM direct-search data.
In section\,\,\ref{hcpv} we scan the model's parameter space that can bring about large $CP$ violation in nonleptonic hyperon decays, taking into account further restrictions from the kaon sector.
Our numerical results illustrate the importance of hyperon and kaon measurements as complementary tools for pinning down new $CP$-violation sources.
In section \ref{concl} we present our conclusions.
A couple of appendices collect additional information pertaining to the hyperon amplitudes.

\section{Darkon model with two Higgs doublets\label{model}}

\subsection{Model description}
Our model of interest is named THDM+D, which is the two-Higgs-doublet model of type III supplemented with a real scalar particle $D$ called darkon and playing the role of the dark matter.
An important feature of the type-III model is that it can offer flavor-changing neutral-Higgs transitions at tree level.

In this scenario the Higgs-quark interactions are described by the Yukawa Lagrangian~\cite{Branco:2011iw}
\begin{align} \label{LYo}
{\cal L}_{\textsc y}^{} & \,=\, -\big(\hat{\textsc y}{}_a^\textsc d\big)_{jk}\, \overline{\textit{\texttt q}_j^{}} \textit{\texttt d}_k^{}\, H_a^{} - \big(\hat{\textsc y}{}_a^\textsc u\big)_{jk}\, \overline{\textit{\texttt q}_j^{}} \textit{\texttt u}_k^{} \tilde H_a^{} \,+\, \rm H.c. \,, &
\end{align}
where \,$a=1,2$\, and \,$j,k=1,2,3$\, are summed over, $\textit{\texttt q}_j^{}$ $(\textit{\texttt d}_k$ and $\textit{\texttt u}_k)$ represent the left-handed quark doublets (right-handed down- and up-type quark fields, respectively), $H_{1,2}$ denote the Higgs doublets, \,$\tilde H_a^{} = i\tau_2^{}H_a^*$,\, with $\tau_2^{}$ being the second Pauli matrix, and $\hat{\textsc y}{}_a^\textsc{d,u}$
are 3$\times$3 matrices for the Yukawa couplings.
In terms of the Higgs components,
\begin{align}
H_a^{} & \,=\, \Bigg(\!\begin{array}{c} h^+_{a} \\ \frac{1}{\sqrt2} \big(v_a^{}+ h_a^0 + i I_a^0\big) \end{array}\!\Bigg) \,, &
\end{align}
where $v_a^{}$ is the vacuum expectation value of $H_a^{}$ satisfying \,$v_1^2+v_2^2=v^2$,\,
with \,$v\simeq246$\,GeV.
The components $h_a^+$, $h_a^0$, and $I_a^0$ are connected to the physical Higgses $h$, $H$, $A$,
and $H^+$ and the Goldstones $w^+$ and $z$ by
\begin{align}
\Bigg(\! \begin{array}{c} h^+_1 \vspace{2pt} \\ h^+_2 \end{array} \!\Bigg) & \,=\, \Bigg(\! \begin{array}{c}c_\beta^{} ~~ {-}s_\beta^{} \\ s_\beta^{} ~~~~ c_\beta^{} \end{array} \!\Bigg) \bigg(\! \begin{array}{c} w^+ \\ H^+ \end{array} \!\bigg) \,, &&
\Bigg(\! \begin{array}{c} I_1^0 \vspace{2pt} \\ I_2^0 \end{array} \!\Bigg) \,=\,
\Bigg(\! \begin{array}{c} c_\beta^{} ~~{-} s_\beta^{} \\ s_\beta^{} ~~~~ c_\beta^{} \end{array} \!\Bigg) \bigg(\! \begin{array}{c} z \\ A \end{array} \!\bigg) \,, ~~~ ~~
\nonumber \\
\Bigg(\! \begin{array}{c} h_1^0 \vspace{3pt} \\ h_2^0 \end{array} \!\Bigg) & \,=\, \Bigg(\! \begin{array}{c} c_\alpha^{} ~~ {-}s_\alpha^{} \\ s_\alpha^{} ~~~~ c_\alpha^{} \end{array} \!\Bigg) \bigg(\! \begin{array}{c} H \\ h \end{array} \!\bigg) \,, &&
c_{\cal X}^{} \,=\, \cos{\cal X} \,, ~~~ s_{\cal X}^{} \,=\, \sin{\cal X} \,,
\end{align}
with  \,$\cos\beta=v_1^{}/v$\, and \,$\sin\beta=v_2^{}/v$.\,

Hereafter, for definiteness we pick \,$\alpha=-\pi/2$\, and \,$\beta=0$,\, implying that $h_1^0$ is $h$ and has no tree-level flavor-changing couplings, \,$h_2^0=-H$, \,$I_1^0=z$, \,$I_2^0=A$, \,$v_1^{}=v$,\, $v_2^{}=0$,\, and only $\hat{\textsc y}{}_1^\textsc{d,u}$ give rise to the quark masses.
After the quark fields have been rotated to their mass eigenstates, the Yukawa interactions of the heavy Higgses become
\begin{align} \label{LY}
{\cal L}_{\textsc y}^{} & \supset\, \tfrac{1}{\sqrt2} \Big[ \big(\texttt Y_2^\textsc d\big)_{jk} \overline{\mathscr D_j^{}} P_R^{} \mathscr D_k^{} + \big(\texttt Y_2^\textsc u\big)_{kj}^* \overline{\mathscr U_j^{}} P_L^{} \mathscr U_k^{} \Big] (H-iA) - \overline{\mathscr U_j^{}} \Big[ \big(\mathscr V\texttt Y_2^\textsc d\big)_{jk} P_R^{} - \big( \texttt Y_2^\textsc u{}^\dagger\mathscr V\big)_{jk} P_L^{} \Big] \mathscr D_k^{} H^+
\nonumber \\ & ~~~ +~ \rm H.c. \,, &
\end{align}
where the transformed Yukawa matrices $\texttt Y_2^\textsc{d,u}$ are generally nondiagonal, \,$\mathscr D_{1,2,3}$ $(=d,s,b)$\, and \,$\mathscr U_{1,2,3}$ $(=u,c,t)$\, refer to the mass eigenstates, \,$P_{L,R}=\big(1\mp\gamma_5^{}\big)/2$,\, and $\mathscr V$ is the Cabibbo-Kobayashi-Maskawa (CKM) matrix.
In the following, we suppose that $H^\pm$, $H$, and $A$ are degenerate in mass and heavier than $h$ and that $\texttt Y_2^\textsc u$ is absent.
We assume additionally that only the first doublet $H_1$ has nonzero Yukawa couplings to SM leptons, endowing mass to the charged ones, implying that these heavy Higgses do not interact directly with the leptons.

The kinetic Lagrangian yields the single Higgs transitions to the $W$ and $Z$ bosons given by
\begin{align} \label{vvh}
{\cal L}_{\rm K}^{} & \,\supset\, \frac{1}{v} \big(2m_W^2 W^{+\kappa}W_\kappa^- + m_Z^2 Z^\kappa Z_\kappa^{} \big) \big[h\,s_{\beta-\alpha}+H\,c_{\beta-\alpha}\big]
\,=\, \frac{h}{v} \big(2m_W^2W^{+\kappa}W_\kappa^-+m_Z^2Z^\kappa Z_\kappa^{}\big)
\end{align}
with \,$\beta-\alpha=\pi/2$,\, and consequently \,$H\to WW,ZZ$\, cannot happen at tree level.
Thus, the aforesaid $\alpha$ and $\beta$ choices simplify things, helping later on make clear the possibility that our model can create sizable hyperon $CP$-violation.
They are also phenomenologically desirable, rendering the lightest Higgs boson, $h$, very SM-like in its couplings to the $W$, $Z$,  and fermions.

In the DM sector, to  ensure the stability of the darkon, $D$, we assume it to be a SM-gauge singlet and introduce a $Z_2$ symmetry under which \,$D\to-D$,\, the other fields being unaltered.
Its renormalizable Lagrangian then takes the form \,${\cal L}_D = \tfrac{1}{2} \partial^\mu D\,\partial_\mu D - {\cal V}_D$,  where~\cite{Bird:2006jd,He:2007tt}
\begin{align} \label{LD2hdmd}
{\cal V}_D^{} & \,=\, \tfrac{1}{2} m_0^2D^2 + \big( \lambda_1^{}H_1^{\dag}H_1^{}
+ \lambda_2^{}H^\dagger_2 H_2^{} + \lambda_3^{} H_1^{\dag}H_2^{}
+ \lambda_3^* H_2^\dagger H_1^{} \big)  D^2 + \tfrac{1}{4} \lambda_D^{}D^4 \,, &
\end{align}
with $m_0^2$ and $\lambda_{1,2,D}$ being real constants due to the hermiticity of ${\cal V}_D$.
Hereafter, we take $\lambda_3$ to be real.\footnote{This can generally be accomplished by absorbing the phase of $\lambda_3^{}$ into $H_1^\dagger H_2^{}$. As a consequence, after the SM fields have been fixed, the phases of the new Yukawa couplings are also fixed.}
After electroweak symmetry breaking, ${\cal L}_D$ contains the darkon's mass
$m_D^{}$ and the $D$-Higgs terms \,$\big({-}\lambda_h^{}h-\lambda_H^{}H\big)D^2v$,\,
but no $DDA$ coupling because $\lambda_3$ is real, where
\begin{align} \label{lambda}
m_D^2 & \,=\, m_0^2 + \big[ \lambda_1^{} \cos^2\!\beta + \lambda_2^{} \sin^2\!\beta
+ \lambda_3^{}\sin(2\beta) \big]v^2 \,=\, m_0^2 + \lambda_1^{} v^2 \,,
\nonumber \\
\lambda_h^{} & \,=\, -\lambda_1^{} \sin\alpha\,\cos\beta + \lambda_2^{} \cos\alpha\, \sin\beta +
\lambda_3^{} \cos(\alpha+\beta) \,=\, \lambda_1^{} \,,
\nonumber \\
\lambda_H^{} & \,=\, \lambda_1^{} \cos\alpha\,\cos\beta + \lambda_2^{} \sin\alpha\,\sin\beta +
\lambda_3^{} \sin(\alpha+\beta) \,=\, -\lambda_3^{} \,. &
\end{align}
Since $\lambda_1$ and $\lambda_3$ are free parameters, we choose $|\lambda_1|$ to be much smaller than $|\lambda_3|$, so that $H$, instead of $h$, is the main mediator between the darkon and SM particles.

In the scalar potential of the model there are also various couplings associated with the interactions of the Higgs doublets' components among themselves, which do not directly pertain to our purposes in what follows.
We assume that all of these parameters already comply with perturbativity, unitarity, vacuum stability, and other requisites to which they may be subject~\cite{Branco:2011iw,He:2016mls}.

\subsection{SMEFT-like framework}

The low-energy processes of interest in this study can be well described within a framework resembling the SM effective field theory (SMEFT)~\cite{Grzadkowski:2010es}.
Thus, with $\texttt Y_2^\textsc u$ being absent, after the heavy Higgs fields have been integrated out, the leading-order SM-gauge-invariant DM-quark operator occurs at dimension six in the effective Lagrangian
\begin{align} \label{eq:DSMEFT}
\mathcal{L}_{D^2 q^2}^{D\tt SMEFT} & \,=\,  \raisebox{3pt}{\small$\displaystyle\sum_{j,k}$}~
C_{qdH D^2}^{jk}\, {\cal Q}_{qdH D^2}^{jk} \,+\, \rm H.c. \,, &
\end{align}
where
\begin{align}
C_{qdHD^2}^{jk} & \,=\, \frac{\lambda_3^{}\mathbb Y_{jk}^{}}{2 m_H^2} \,,
~~~~~ \mathbb Y \,\equiv\, \texttt Y_2^\textsc d \,, & {\cal Q}_{qdH D^2}^{jk}  & \,=\, \overline{\textit{\texttt q}}_j^{} \textit{\texttt d}_k^{} H_1^{} D^2 \,. &
\end{align}
The tree-level matching result for dim-6 operators consisting of quark fields is expressible as
\begin{align} \label{L4qsmeft}
{\cal L}_{4q}^{\tt SMEFT} & \,=\, \raisebox{3pt}{\small$\displaystyle\sum_{j,k,l,o}$}~
 C_{qddq}^{jklo}\, {\cal Q}_{qddq}^{jklo} \,, &
\end{align}
where \,$j,k,l,o=1,2,3$,
\begin{align}
C_{qddq}^{jklo} \,=\, \frac{\mathbb Y_{jk}^{} \mathbb Y_{lo}^{\dagger}}{m_H^2} \,,  &&
{\cal Q}_{qddq}^{jklo} \,=\, \big(\overline {\textit{\texttt q}}{}_j \textit{\texttt d}_k^{}\big) \big(\overline{\textit{\texttt d}}_l^{} \textit{\texttt q}_o^{}\big) \,, ~~~~~ &
\end{align}
with \,$m_{H^+}^{}=m_A^{}=m_H^{}$.

\subsection{LEFT-like framework}

To treat processes at a scale of a few GeV, the above SMEFT-like Lagrangian has to be matched onto a framework similar to the low-energy effective field theory (LEFT)~\cite{Jenkins:2017jig} by taking into account electroweak symmetry breaking and integrating out the SM heavy states ($t,h,W,Z$).
After the Higgs' vacuum expectation value develops, the darkon-quark terms in eq.\,(\ref{eq:DSMEFT})  become, in the mass basis of the down-type quarks,
\begin{align} \label{eq:DLEFT}
\mathcal{L}_{q^2 D^2}^{D\tt LEFT} & \,=\, \frac{D^2}{2}\,  \raisebox{3pt}{\small$\displaystyle\sum_{j,k}$}\, \Big( C_{dD}^{S,jk}\, \overline{\mathscr D_j^{}} \mathscr D_k^{} + C_{dD}^{P,jk}\, \overline{\mathscr D_j^{}} i\gamma_5^{} \mathscr D_k^{} \Big) \,, &
\end{align}
where
\begin{align}
\label{eq:leftWCmatching}
C_{dD}^{S,jk} & \,=\, \lambda_3^{} v~ \frac{\mathbb Y_{jk}^{} + \mathbb Y_{kj}^{*}}{\sqrt2\, m_H^2} \,, &
i C_{d D}^{P,jk} & \,=\, \lambda_3^{} v~ \frac{\mathbb Y_{jk}^{} - \mathbb Y_{kj}^{*}}{\sqrt2\, m_H^2} \,, & C_{dD}^{S(P),kj} & \,=\, C_{dD}^{S(P),jk*} .
\end{align}
From eq.\,(\ref{L4qsmeft}), in the LEFT the matching result for the four-quark operators is given by
\begin{equation} \label{L4qleft}
{\cal L}_{4q}^{\tt LEFT} \,=\, \raisebox{3pt}{\small$\displaystyle\sum_{j,k,l,o}$} \Big( C_{uddu}^{jklo}\, {\cal Q}_{uddu}^{jklo} + C_{dddd}^{jklo}\, {\cal Q}_{dddd}^{jklo} \Big) \,,
\end{equation}
and their matching onto the SMEFT at electroweak scale is
\begin{align}
C_{uddu}^{jklo} & \,=\, \mathscr V_{jx}^{} \mathscr V_{oy}^*\, C_{qddq}^{klxy} \,=\, \mathscr V_{jx}^{} \mathscr V_{oy}^*\, \frac{\mathbb Y_{xk}^{} \mathbb Y_{yl}^{*}}{m_H^2} \,, & {\cal Q}_{uddu}^{jklo} & \,=\, \big( \overline{u_{jL}} d_{kR}^{} \big) \big( \overline{d_{lR}} u_{oL}^{} \big) \,, ~~~~~
\nonumber \\
C_{dddd}^{jklo} & \,=\,  C_{qddq}^{jklo} \,=\, \frac{\mathbb Y_{jk}^{} \mathbb Y_{ol}^{*}}{m_H^2} \,, & {\cal Q}_{dddd}^{jklo} & \,=\, \big( \overline{d_{jL}} d_{kR}^{} \big) \big( \overline{d_{lR}} d_{oL}^{} \big) \,. \end{align}

\section{Phenomenology of  \boldmath$B^+\to K^+\rm{+}invisible$ excess\label{b2sdd}}

The $B\to KDD$ and $B\to K^*DD$ modes are induced, respectively, by the $\overline bsD^2$ and $\overline bi\gamma_5^{}sD^2$ terms in eq.\,(\ref{eq:DLEFT}).
The corresponding differential rates have been calculated in ref.\,\cite{He:2022ljo} to be
\begin{align} \label{G'B2KDD}
\frac{d\Gamma_{B\to KDD}}{d\hat\varsigma} & \,=\, \frac{\lambda^{1/2}\big(m_B^2, m_K^2,\hat\varsigma\big)\,  \kappa^{1/2}\big(m_D^2,\hat\varsigma\big)}{512 \pi^3 m_B^3} \Bigg(\frac{m_B^2 - m_K^2}{m_b^{}-m_s^{}}\Bigg)\!\raisebox{1em}{$^2$} \Big|C_{dD}^{S,bs} f_0^{BK}(\hat\varsigma)\Big|^2 ,
\nonumber \\
\frac{d\Gamma_{B\to K^*DD}}{d\hat\varsigma} & \,=\, \frac{\lambda^{3/2}(m_B^2, m_{K^*}^2, \hat\varsigma)\, \kappa^{1/2}(m_D^2,\hat\varsigma)}{512 \pi^3 m_B^3\, (m_b + m_s)^2} \Big|C_{dD}^{P,bs} A_0^{BK^*}(\hat\varsigma)\Big|^2 , &
\end{align}
where $\hat\varsigma$ denotes the $DD$ pair's squared invariant-mass, $f_0^{BK}$ and $A_0^{BK^*}$ represent form factors depending on $\hat\varsigma$,
\begin{align} &
\lambda(x,y,z) \,=\, (x-y-z)^2 - 4yz \,, & & \kappa(m_D^2,\hat\varsigma) \,=\, 1 - 4m_D^2/\hat\varsigma \,,
\nonumber \\ &
C_{dD}^{S,bs} \,= \hat\eta_b^{} \lambda_3^{} v\, \frac{\mathbb Y_{bs}^{} + \mathbb Y_{sb}^*}{\sqrt2\, m_H^2} \,, & &
i C_{d D}^{P,bs} \,= \hat\eta_b^{} \lambda_3^{} v\, \frac{\mathbb Y_{bs}^{} - \mathbb Y_{sb}^*}{\sqrt2\, m_H^2} \,. & \label{CSCP}
\end{align}
In numerical work, we adopt the $f_0^{BK}$ and $A_0^{BK^*}$ results of ref.\,\cite{Gubernari:2023puw}.
Moreover, in eq.\,(\ref{CSCP}) we have included the constant
\,$\hat\eta_b^{}=1.8$\, due to QCD renormalization group evolution from the NP scale of 1~TeV to the relevant hadronic scale, which is taken to be the $B$-meson mass.
Although $\mathbb Y_{bs}$ and $\mathbb Y_{sb}$ can generally influence $B_s$-$\bar B_s$ mixing, for \,$m_A=m_H$\, the effect is negligible if one of them is much less than the other~\cite{Crivellin:2017upt}.
The findings of ref.\,\cite{He:2024iju} imply that to explain the Belle II excess in the THDM+D with \,$m_D^{}\in[0.1,1]$\,\,GeV\, requires $|C_{dD}^{S,sb}|$ to be in the range \,$[3,7]/\big(10^5\, \rm TeV\big)$,\, which for \,$m_H^{}=1$\,\,TeV\, translates into \,$|\lambda_3 (\mathbb Y_{bs}^{} + \mathbb Y_{sb}^*)|\sim  [1,2.6]\times 10^{-4}$.

Analogously the interactions  in eq.\,(\ref{eq:DLEFT}) can bring about the quark transition \,$s\to dDD$\, which causes the kaon decay \,$\bar K\to\pi DD$\, if the darkon mass is sufficiently small.
It turns out, however, that it is not feasible for the desired model parameter space with such a light darkon to reproduce the correct DM relic abundance without conflicting with DM direct-search data.
Consequently, in the rest of this paper we ignore the \,$m_D\le(m_K-m_\pi)/2$\, region.

\section{DM relic density and direct search constraints\label{DMsector}}

\subsection{Relic density\label{dmrelic}}

We restrict our analysis to DM masses below the $K^*$-meson threshold, which are also favored by the explanation for the \,$B^+\to K^+{+}\rm invisible$\, anomaly, i.e., $(m_K-m_\pi)/2<m_D \lesssim 900\,\rm MeV$.
In this range, the relic abundance is produced mainly via darkon  annihilation into light pseudoscalar mesons at a temperature of tens of MeV.
Thus, it is appropriate to employ chiral perturbation theory to describe their interactions with the darkon.
From eq.\,(\ref{eq:DLEFT}), the pertinent leading effective Lagrangian is then \cite{He:2024iju}
\begin{align} \begin{array}[b]{l} \displaystyle
{\cal L}_{D P}^{} \,\supset\, \frac{B_0 D^2}{2} \Bigg\{ \frac{C_{d D}^{S,dd}}{2}  \big( 2 \pi^+ \pi^- + \pi^0 \pi^0 \big) +  C_{d D}^{S,ss}  K^+ K^- + \big(C_{d D}^{S,dd}+ C_{d D}^{S,ss}\big) K^0 \bar K^0 - \frac{C_{d D}^{S,dd}\pi^0\eta_8^{}}{\sqrt3}
\\ \displaystyle \hspace{7em} +~ \big(C_{dD}^{S,dd} +  4 C_{d D}^{S,ss}\big) \frac{\eta_8^2}{6} + \Bigg[ C_{d D}^{S,ds} \Bigg(\pi^+ K^- - \frac{\pi^0\bar K^0}{\sqrt2}
- \frac{\eta_8^{}\bar K^0}{\sqrt6} \Bigg)
+ {\rm H.c.} \Bigg] \Bigg\} \,, \end{array} \label{eq:LDP}
\end{align}
where  \,$B_0=m_K^2/(m_s+m_d)\sim2$ GeV,\,
\begin{align} \label{CdDS}
C_{dD}^{S,dd(ss)} & \,=\, \hat\eta_s^{} \lambda_3^{} v\, \frac{\sqrt2\, {\rm Re}\, \mathbb Y_{dd(ss)}}{m_H^2} \,, & C_{dD}^{S,ds} & \,=\,
\hat\eta_s^{} \lambda_3^{} v~ \frac{\mathbb Y_{ds}^{} + \mathbb Y_{sd}^{*}}{\sqrt2\, m_H^2} \,, & C_{u D}^{S,uu} & \,=\, 0 \,,
\end{align}
with \,$\hat \eta_s^{}=2.3$\, having been added due to QCD running from 1 TeV to the hadronic scale of 1\,\,GeV.
We can rewrite each term in eq.\,(\ref{eq:LDP}) as
\begin{equation}
\mathcal{L}_{DP}^{} \,\supset\, \frac{\kappa_{ab}^{}}{2} D^2 \textsl{\texttt M}_a \textsl{\texttt M}_b \,,
\end{equation}
where $\textsl{\texttt M}_{a,b}$ stand for pseudoscalar mesons and $\kappa_{ab}$ can be easily read off from eq.\,(\ref{eq:LDP}).
It follows that the cross section of darkon annihilation into $\textsl{\texttt M}_a \textsl{\texttt M}_b$ is
\begin{equation}
    \sigma(DD\to \textsl{\texttt M}_a \textsl{\texttt M}_b) \,=\, (1 + \delta_{ab}) \frac{|\kappa_{ab}|^2}{16\pi\hat\varsigma} \frac{\lambda^{1/2}\big(\hat\varsigma,m_a^2,m_b^2\big)}{\lambda^{1/2}\big(\hat \varsigma,m_D^2,m_D^2\big)} \,.
\end{equation}
This leads to the thermally averaged cross-section  \cite{Gondolo:1990dk}
\begin{align}
\langle \sigma v\rangle & \,=\, (1 + \delta_{ab}) \frac{|\kappa_{ab}|^2\, \tilde\eta(x,z_a,z_b)}{64\pi m_D^2} \,,
\nonumber \\
\tilde\eta(x,z_a,z_b) & \,\equiv\, \frac{4 x}{K_2^2(x)} \int_0^\infty  \frac{d\tilde\epsilon\,\sqrt{\tilde\epsilon}}{1+\tilde\epsilon}\, \lambda^{1/2} \big(1+\tilde\epsilon,z_a/4,z_b/4\big)\, K_1^{} \big(2 x\sqrt{1+\tilde\epsilon}\big) \,, &
\end{align}
where \,$x=m_D/T$\, with $T$ being the temperature, $K_{1,2}$ are modified Bessel functions, $z_i=m_i^2/m_D^2$, and \,$\tilde\epsilon \equiv \hat\varsigma/\big(4 m_D^2\big) - 1$.
This cross section must be
\,$\langle\sigma v \rangle \simeq 2.2\times 10^{-9}\rm\, GeV^{-2}$~\cite{Steigman:2012nb} if the slight variation between the kaon mass $m_K$ and 1 GeV is neglected.
A more precise evaluation  can be done with the aid of figures\,\,4 and 5 in ref.~\cite{Steigman:2012nb}.
Specifically, to extract the corresponding values of $|C_{d D}^{S,jk}|$, we digitize those two figures to obtain the matching point \,$x=x_*$\, of the two regimes of the freeze-out process as a function of the DM mass and the thermally averaged cross-section at $x_*$ needed to find the correct relic abundance.\footnote{The evolution of the DM abundance can be separated into two regimes (``early'' and ``late'') in which the evolution equation
can be solved analytically by different approximations \cite{Steigman:2012nb}. The solutions are then joined at an intermediate matching point called $x_*$.
Further details can be found in ref.\,\cite{Steigman:2012nb}.}

\begin{figure}[t!] \bigskip \centering
\includegraphics[width=0.48\linewidth]{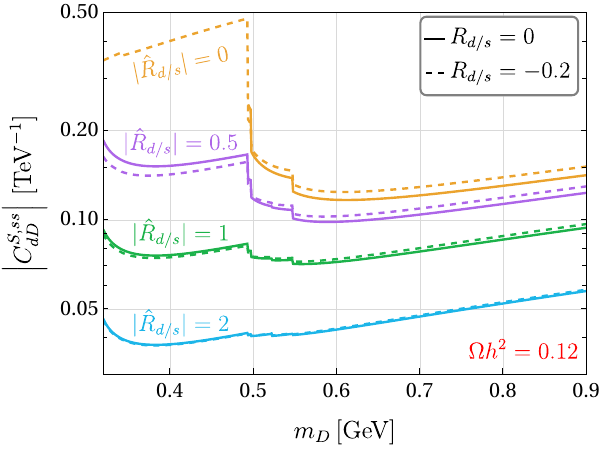} ~ \includegraphics[width=0.48\linewidth]{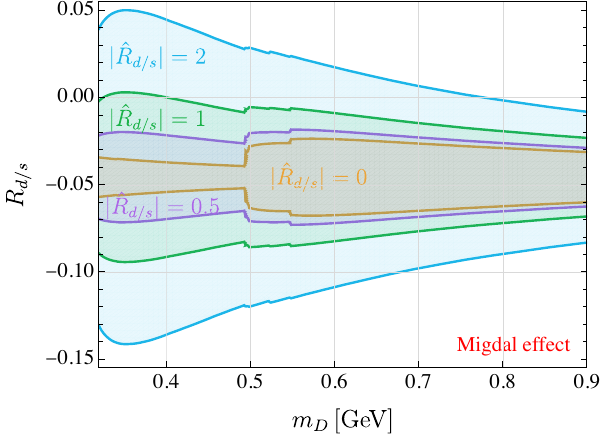}\vspace{-1ex}
\caption{Left: the values of $\big|C_{dD}^{S,ss}\big|$ versus the darkon mass $m_D$ that are compatible with the observed DM relic density for \,$R_{d/s}=0$ (solid curves), $-0.2$ (dashed cures) and \,$\big|\hat R_{d/s}\big|=0$ (orange), 0.5 (purple), 1 (green), 2 (cyan).
Right: the $m_D$-$R_{d/s}$ regions for the same $\big|\hat R_{d/s}\big|$ choices that are allowed by the DM direct search result of PandaX-4T \cite{PandaX:2023xgl} incorporating the Migdal effect.
Note that, for the right panel, $\big|C_{dD}^{S,ss}\big|$ is fixed by the DM relic density requirement shown as solid curves on the left panel.
} \label{fig:dm}
\end{figure}

For numerical convenience, we define
\begin{align} \label{Rd/s}
R_{d/s}^{} & \,=\, \frac{C_{dD}^{S,dd}}{C_{dD}^{S,ss}} \,=\, \frac{{\rm Re}\, \mathbb Y_{dd}}{{\rm Re}\, \mathbb Y_{ss}} \,, & \hat R_{d/s}^{} & \,=\, \frac{C_{dD}^{S,ds}}{C_{dD}^{S,ss}} \,=\, \frac{\mathbb Y_{ds}^{} + \mathbb Y_{sd}^{*}}{2\, {\rm Re}\, \mathbb Y_{ss}} \,. ~~~~~
\end{align}
On the left panel of figure \ref{fig:dm} we present the $|C_{d D}^{S,ss}|$ ranges that can translate into the correct relic density as a function of the darkon mass $m_D$ for different choices of $R_{d/s}$ and $|\hat R_{d/s}|$.
A comparison of this figure and the $|C_{dD}^{S,sb}|$ range of \,$[3,7]/\big(10^5\, \rm TeV\big)$,\, which as discussed in section~\ref{b2sdd} can account for the Belle II anomaly, clearly demonstrates the necessity for a substantial $\big(${\small$\sim$\,}$10^3\big)$ hierarchy between $\mathbb Y_{sb}$ and $\mathbb Y_{ss}$.

\subsection{DM direct searches\label{dmdd}}

It is well known that in DM direct searches relying on the usual nuclear recoil for signal detection the sensitivity to sub-GeV DM decreases considerably because of detection threshold issues.
Such experiments therefore would not yield any useful restraints on our parameter space of interest.
It has been suggested, however, that the Migdal effect can be used to bypass this kinematic obstacle and probe the light-DM region.
This proposal has been implemented in the liquid xenon and argon experiments including XENON1T \cite{XENON:2019zpr}, DarkSide50 \cite{DarkSide:2022dhx}, LZ \cite{LZ:2023poo}, and PandaX-4T \cite{PandaX:2023xgl}, resulting in meaningful restraints on the spin-independent DM-nucleon cross-section.
To apply them to our DM scenario, we calculate the corresponding darkon-nucleon cross section in the limit of zero momentum transfer.
The operators
\,${\cal Q}_{d D}^{S,ss}=\frac{1}{2} \overline s s D^2$\, and \,${\cal Q}_{d D}^{S,dd}=\frac{1}{2} \overline d d D^2$\, associated with $C_{dD}^{S,ss}$ and $C_{dD}^{S,dd}$ can induce the usual spin-independent DM-nucleon nonrelativistic operator \,${\cal O}_1^N \equiv 1_\phi 1_N$\, for \,$N=$ proton ($p$) or neutron ($n$) with the coefficient
\begin{equation}
c_1^N \,=\, \frac{2 m_N^2}{m_s} f_{T_s}^{(N)} C_{d D}^{S,ss} + \frac{2 m_N^2}{m_d} f_{T_d}^{(N)} C_{d D}^{S,dd} \,=\, \frac{2 m_N^2}{m_s} \big( 1 + r_N R_{d/s} \big) f_{T_s}^{(N)} C_{d D}^{S,ss} \,,
\end{equation}
where \,$r_N^{} \equiv f_{T_d}^{(N)} m_s^{}/\big[f_{T_s}^{(N)} m_d^{}\big]$\, and $f_{T_q}^{(N)}$ is a nucleon form factor in the scalar quark current at zero momentum transfer.
Numerically, we employ \,$f_{T_s}^{(N)}=0.044$, \,$f_{T_d}^{(p)}=0.038$, \,$f_{T_d}^{(n)}=0.056$  \cite{DelNobile:2021wmp}, \,$m_d=6.2$\,\,MeV, and \,$m_s=123$\,\,MeV,\, leading to \,$r_p \simeq 17$\, and \,$r_n \simeq 25$.\,
A nonzero value of  $R_{d/s}$ implies \,$c_1^p\neq c_1^n$,\, which corresponds to the isospin-violating DM scenario.

Incorporating these isospin-violating couplings, the corresponding effective darkon-nucleon cross-section is computed to be \cite{He:2024iju}
\begin{equation}
\sigma_{D N}^{} \,=\, {\mu_{D N}^2  \over 4\pi m_D^2} \left| {m_N\over m_s} f_{T_s}^{(N)}{\cal C}_{d D}^{S,ss} \right|^2
\left|
 \left( 1 + r_p R_{d/s}\right) {Z \over A}
 +
\left( 1 + r_n R_{d/s} \right){A-Z\over A}
\right|^2 \,,
\end{equation}
where $\mu_{DN}^{}$ denotes the reduced mass of the darkon-nucleon system and $A$ $(Z)$ the atomic (mass) number of the target nucleus.
The DarkSide experiment utilizes a pure argon detector with only one isotope, $\rm{}^{40}Ar$.
Xenon-target experiments (XENON1T, LZ, and PandaX-4T) involve several isotopes with abundances summarized in the table\,\,1 of ref.\,\cite{DelNobile:2021wmp}.
Accordingly, we need to use the weighted average \,$\sum_i \hat\xi_i\,\sigma_{DN}^i$, where the sum runs over all isotopes $i$ with abundance $\hat\xi_i$.
In the right panel of figure\,\,\ref{fig:dm} we have graphed the allowed regions in the $m_D$-$R_{d/s}$ plane for a few selections of $\hat R_{d/s}$ and $|\hat R_{d/s}|$.
As clearly shown in the plot, the region around \,$R_{d/s}\simeq-0.04$\, is consistently permitted, regardless of the $|\hat R_{d/s}|$ choice.
This is due to the fact that this $R_{d/s}$ value results in a significant cancellation between the proton and neutron contributions to the effective darkon-nucleon cross-section.

\section{Hyperon \boldmath$CP$ violation\label{hcpv}}

Nonleptonic hyperon decays are affected by new four-quark interactions mediated at tree level by the heavy Higgses and  described by the effective Lagrangian in eq.\,\,(\ref{L4qleft}),
\begin{align} \label{L4q}
{\cal L}_{4q}^{\tt LEFT} & \,\supset\, {\cal C}_u^{} {\cal Q}_u^{} + {\cal C}_+^{} {\cal Q}_+^{} + {\cal C}_-^{} {\cal Q}_-^{} \,+\, \rm H.c. \,, ~~~ ~~~~
\end{align}
where
\begin{align} \label{C-+u}
{\cal C}_u^{} & \,=\, V_{us}^{} \mathbb Y_{ss}^{}\, \frac{ V_{ud}^* \mathbb Y_{dd}^* + V_{us}^* \mathbb Y_{sd}^* }{m_H^2} \,, & {\cal Q}_u^{} & \,=\, \overline{u_L^{}}s_R^{}\, \overline{d_R^{}}u_L^{} \,,
\nonumber \\
{\cal C}_\pm^{} & \,=\, \mathbb Y_{sd}^*\, \frac{\mathbb Y_{dd}^{}\pm\mathbb Y_{ss}^{}}{2 m_H^2} \,\equiv\, \frac{{\cal C}_d\pm{\cal C}_s}{2} \,, & {\cal Q}_\pm^{} & \,=\, \overline{d_R^{}}s_L^{}\, \big( \overline{d_L^{}}d_R^{} \pm \overline{s_L^{}}s_R^{} \big) \,. &
\end{align}
We have set \,$\mathbb Y_{ds}=0$\, in order to remove the short-distance effects of \,$\mathbb Y_{ds,sd}$\, on kaon mixing.\footnote{The contribution at short distance to kaon mixing is vanishing if \,$m_A=m_H$,\, which we already picked, and either $\mathbb Y_{ds}$ or $\mathbb Y_{sd}$ is zero~\cite{Crivellin:2017upt}.}

To deal with the impact of eq.\,(\ref{L4q}) on $CP$ violation in nonleptonic hyperon decay, we employ a chiral-Lagrangian approach.
In this context, the relevant baryon and meson fields are those of the lightest octets collected into the matrices
\begin{align} \label{BSx}
B & \,= \left(\!\begin{array}{ccc} \displaystyle \frac{\Lambda^{\vphantom{|^|}}}{\sqrt6} + \frac{\Sigma^0}{\sqrt2} & \Sigma^+ & p \\ \Sigma^- & \displaystyle \frac{\Lambda}{\sqrt6} - \frac{\Sigma^0}{\sqrt2} & n \\ \Xi^- & \Xi^0 & \displaystyle -\frac{\sqrt2}{\sqrt3_{\vphantom{|}}}\,\Lambda \end{array}\!\right) , &
\varphi & \,= \left(\!\begin{array}{ccc} \displaystyle \frac{\eta_8^{\vphantom{|}}}{\sqrt3} + \pi^0
& \sqrt2\,\pi^+ & \sqrt2\,K^+ \\ \sqrt2\, \pi^- & \displaystyle \frac{\eta_8^{}}{\sqrt3} - \pi^0 & \sqrt2\, K^0 \\ \sqrt2\, K^- &
\sqrt2~\overline{\!K}{}^0 & \displaystyle \frac{-2\eta_8^{}}{\sqrt3_{\vphantom{|}}}
\end{array}\!\right) , ~
\nonumber \\
\overline{\!B} & \,=\, B^\dagger \gamma_0^{} \,, & \Sigma & \,=\,  \xi^2 \,=\, e^{i\varphi/f_\pi^{}} \,,
\end{align}
where $f_\pi^{}$ is the pion decay constant.
They enter the leading-order strong chiral Lagrangian~\cite{Bijnens:1985kj}
\begin{align} \label{Lstrong}
{\cal L}_{\rm s}^{} \,\supset\, {\rm Tr} & \Big[ \overline B i\gamma^\kappa \Big(
\partial_\kappa^{} B + \tfrac{1}{2} \Big[ \xi\,\partial_\kappa^{} \xi^\dagger + \xi^\dagger \partial_\kappa^{} \xi, B \Big] \Big) + \overline B \gamma^\kappa \gamma_5^{} \big( {\cal D}\, \big\{{\cal A}_\kappa^{}, B\big\}
+ {\cal F}\, \big[{\cal A}_\kappa^{}, B\big] \big)
\nonumber \\ & +\, b_D^{}\, \overline B \big\{M_+, B\big\} + b_F^{}\, \overline B \big[M_+,B\big] \Big] \,, &
\end{align}
where $\cal D$ and $\cal F$ $(b_D$ and $b_F)$ are constants with values extractable from baryon semileptonic decays (the octet baryons' masses),  \,${\cal A}^\kappa = \tfrac{i}{2} \big( \xi\,\partial^\kappa \xi^\dagger - \xi^\dagger \partial^\kappa \xi \big)$,\, and \,$M_+ = \xi^\dagger M_q^{} \xi^\dagger + \xi M_q^\dagger \xi$,\,  with \,$M_q = {\rm diag}\big(\hat m,\hat m,m_s^{}\big)$\, and \,$\hat m = (m_u+m_d)/2$.\,
Under the chiral-symmetry group \,${\rm SU}(3)_L\times{\rm SU}(3)_R$\, the matrices in eq.\,(\ref{BSx})
transform as
\begin{align}
B & \,\to\, \hat UB\hat U^\dagger \,, &
\,\overline{\!B} & \,\to\, \hat U\,\overline{\!B}\hat U^\dagger \,, &
\Sigma & \,\to\, \hat L\Sigma\hat R^\dagger \,, &
\xi & \,\to\, \hat L\xi\hat U^\dagger \,=\, \hat U\xi\hat R^\dagger \,, ~~~
\end{align}
where \,$\hat U\in{\rm SU}(3)$\, is implicitly defined by the $\xi$ equation, \,$\hat L\in{\rm SU}(3)_L$,\, and \,$\hat R\in{\rm SU}(3)_R$.\,

The chiral realization of the operators in eq.\,(\ref{L4q}) must share their corresponding symmetry properties.
In this regard, we observe that ${\cal Q}_u$ is invariant under charge conjugation (\textsl{\texttt C}) followed by a parity operation (\textsl{\texttt P}) and switching the $d$ and $s$ quarks (\textsl{\texttt S}), while ${\cal Q}_{+(-)}$ is \textsl{\texttt{CS}} even (odd).
Thus, we have the correspondences at leading order
\begin{align} \label{Qu}
{\cal Q}_u^{} \Leftrightarrow\, {\cal O}_u^{} \,\equiv & ~ \hat{\textsc d}_u^{}
\Big[ \big( \xi^\dagger \big\{ B, \overline B \big\} \xi^\dagger \big)_{31} \Sigma_{12}^{}
+ \Sigma_{31}^\dagger \big( \xi \big\{ B, \overline B \big\} \xi \big)_{12} \Big]
\nonumber \\ & +\,
\hat{\textsc f}_u^{} \Big[ \big( \xi^\dagger \big[ B, \overline B \big] \xi^\dagger \big)_{31}
\Sigma_{12}^{}+\Sigma_{31}^\dagger \big( \xi \big[ B, \overline B \big] \xi \big)_{12}\Big]
\nonumber \\ & +\, \hat{\textsc g}_u^{}\, \big( \xi^\dagger \overline B \xi^\dagger \big)_{31} (\xi B\xi)_{12}^{}
+ \hat{\textsc g}{}_u'\, \big(\xi \overline B \xi\big)_{12} \big( \xi^\dagger B \xi^\dagger \big)_{31}
+ \hat{\textsc h}_u^{}\, \Sigma_{31}^\dagger \Sigma_{12}^{} \,,
\\ \label{Qpm} \raisebox{4ex}{}
{\cal Q}_\pm^{} \Leftrightarrow\, {\cal O}_\pm^{} \,\equiv & ~ \Big[  \hat{\textsc d}_\pm^{}\, \big( \xi \big\{ B, \overline B \big\} \xi \big)_{32}
+ \hat{\textsc f}_\pm^{}\, \big( \xi \big[ B, \overline B \big] \xi \big)_{32} \Big] \big( \Sigma_{22}^\dagger\pm\Sigma_{33}^\dagger \big)
\nonumber \\ & +\, \Sigma_{32}^{}\, \Big[ \hat{\textsc d}{}_\pm'\,
\big( \xi^\dagger \big\{ B, \overline B \big\} \xi^\dagger \big)_{22} + \hat{\textsc f}{}_\pm'\,
\big( \xi^\dagger \big[ B, \overline B \big] \xi^\dagger \big)_{22} \pm (2{\,\to\,}3) \Big]
\nonumber \\ & +\,
\hat{\textsc g}_\pm^{} \Big\{ \big(\xi \overline B \xi\big)_{32} \Big[ \big( \xi^\dagger B \xi^\dagger \big)_{22} \pm  (2{\,\to\,}3) \Big] + \Big[ \big( \xi^\dagger \overline B \xi^\dagger \big)_{22}
\pm  (2{\,\to\,}3) \Big] \big(\xi B\xi\big)_{32} \Big\}
\nonumber \\ & +\, \hat{\textsc h}_\pm^{}\, \Sigma_{32}^{} \big( \Sigma_{22}^\dagger\pm\Sigma_{33}^\dagger \big) \,, &
\end{align}
where $\hat{\textsc d}_{u,\pm}$, $\hat{\textsc f}_{u,\pm}$, $\hat{\textsc g}_{u,\pm}$,
$\hat{\textsc g}{}_u'$, and $\hat{\textsc h}_{u,\pm}$ are constants.
The chiral Lagrangian from eq.\,(\ref{L4q}) is then
\begin{align} \label{Lchi} &
{\cal L}_\chi^{\rm new} \,=~ \tilde{\textsc c}_u^{}\, {\cal O}_u^{} + \tilde{\textsc c}_+^{}\, {\cal O}_+^{} + \tilde{\textsc c}_-^{}\, {\cal O}_-^{} \,+\, \rm H.c. \,, &
\end{align}
where
\begin{align} \label{etas}
\tilde{\textsc c}_u^{} & \,=\, \hat\eta_u^{}\, {\cal C}_u^{} \,, ~~~ ~~ \hat\eta_u^{} \,=\, 4.7 \,, &
\tilde{\textsc c}_\pm^{} & \,=\, \tfrac{1}{2} \hat\eta_d^{}\,  \big({\cal C}_d^{} \pm {\cal C}_s^{} \big) \,, ~~~ ~~ \hat\eta_d^{}  \,=\, 4.8 \,, ~~~
\end{align}
with the parameters $\hat\eta_{u,d}^{}$ appearing due to QCD renormalization group evolution from 1\,\,TeV to 1\,\,GeV and evaluated with the {\tt flavio} code \cite{Straub:2018kue}.

\subsection{Hyperon decays\label{hyperons}}

\begin{figure}[t!] \centering
\includegraphics[trim=37mm 227mm 38mm 38mm,clip,width=0.9\textwidth]{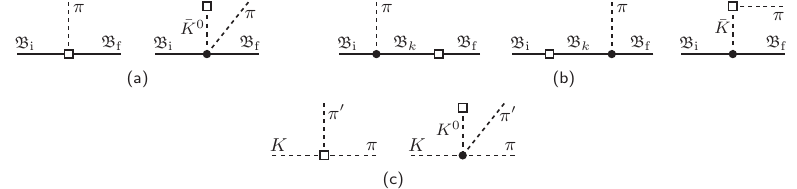}
\vspace{-1ex}\caption{Leading-order diagrams for the new contributions to two-body nonleptonic (a) {\tt S}- and (b) {\tt P}-wave hyperon and (c) kaon decays. Each hollow square (thick dot) symbolizes a coupling from ${\cal L}_\chi^{\rm new}$ in eq.\,(\ref{Lchi}) [${\cal L}_{\rm s}$ in eq.\,(\ref{Lstrong})] and a solid (dashed) line represents a baryon (meson).} \label{diagrams}
\end{figure}

The amplitude for a spin-1/2 baryon $\mathfrak B_{\rm i}$ decaying into another one
$\mathfrak B_{\rm f}$ and a pseudoscalar meson \textsl{\texttt M} has the general form
\begin{align} \label{m_BB'pi}
i {\cal M}_{\mathfrak B_{\rm i}^{}\to\mathfrak  B_{\rm f}^{}\textsl{\texttt M}}^{} & \,=\, i \langle\mathfrak B_{\rm f\,} \textsl{\texttt M}| {\cal H}_{\rm w}^{}|\mathfrak B_{\rm i}\rangle \,=\, \bar u_{\rm f}^{}\, \big(\mathbb A-\gamma_5^{} \mathbb B\big) \, u_{\rm i}^{} \,, &
\end{align}
where $\mathbb A$ and $\mathbb B$ are constants that belong to, respectively, the {\tt S}-wave and {\tt P}-wave
components of the amplitude.
The effect of eq.\,(\ref{Lchi}) on $\mathbb A$ at leading order is calculated from the contact and tadpole diagrams displayed in figure\,\,\ref{diagrams}(a).
Thus, focusing on the $\Lambda$ and $\Xi$ modes, which have so far been the most measured among nonleptonic hyperon decays~\cite{ParticleDataGroup:2024cfk}, we arrive at the $\mathbb A$ expressions for \,$\Lambda\to p\pi^-,n\pi^0$\, and \,$\Xi^{-,0}\to\Lambda\pi^{-,0}$\, in eq.\,(\ref{Aamplitudes}) in appendix\,\,\ref{ABnew}.
The corresponding contributions to $\mathbb B$ come from baryon- and kaon-pole diagrams as drawn in figure\,\,\ref{diagrams}(b).
Accordingly, we get the $\mathbb B$ formulas in eq.\,(\ref{Bamplitudes}).

In general, these transitions change isospin by \,$\Delta I=1/2$ or 3/2,\, and the \,$\textsl{\texttt L}=\mathbb A,\mathbb B$\, amplitudes can be decomposed into their $\Delta I$ components as
\begin{align} \label{MLX}
\textsl{\texttt L}_{\Lambda\to p\pi^-}^{} & \,=\, -\sqrt2~
\textsl{\texttt L}_{1/2}^\Lambda\,
{\rm exp}\big(i\delta_{1\textsl{\texttt L}}^\Lambda+i\xi_{1\textsl{\texttt L}}^\Lambda\big)
+ \textsl{\texttt L}_{3/2}^\Lambda\,
{\rm exp}\big(i\delta_{3\textsl{\texttt L}}^\Lambda+i\xi_{3\textsl{\texttt L}}^\Lambda\big) \,, &
\nonumber \\
\textsl{\texttt L}_{\Lambda\to n\pi^0}^{} & \,=\, \textsl{\texttt L}_{1/2}^\Lambda\,
{\rm exp}\big(i\delta_{1\textsl{\texttt L}}^\Lambda+i\xi_{1\textsl{\texttt L}}^\Lambda\big) + \sqrt2~
\textsl{\texttt L}_{3/2}^\Lambda\,
{\rm exp}\big(i\delta_{3\textsl{\texttt L}}^\Lambda+i\xi_{3\textsl{\texttt L}}^\Lambda\big) \,,
\nonumber \\
\textsl{\texttt L}_{\Xi^-\to\Lambda\pi^-}^{} & \,=\, \Big[ \sqrt2~ \textsl{\texttt L}_{1/2}^\Xi\, {\rm exp}\big(i\xi_{1\textsl{\texttt L}}^\Xi\big) + \textsl{\texttt L}_{3/2}^\Xi\,
{\rm exp}\big(i\xi_{3\textsl{\texttt L}}^\Xi\big) \Big]\, {\rm exp} \big(i\delta_{1\textsl{\texttt L}}^\Xi\big) \,,
\nonumber \\
\textsl{\texttt L}_{\Xi^0\to\Lambda\pi^0}^{} & \,=\, \Big[ {-} \textsl{\texttt L}_{1/2}^\Xi\, {\rm exp} \big(i\xi_{1\textsl{\texttt L}}^\Xi\big) + \sqrt2~ \textsl{\texttt L}_{3/2}^\Xi\, {\rm exp} \big(i\xi_{3\textsl{\texttt L}}^\Xi\big) \Big]\, {\rm exp} \big(i\delta_{1\textsl{\texttt L}}^\Xi\big) \,,
\end{align}
where the \,$\Delta I=1/2$\, parts are empirically known to be dominant and the $\delta$s and $\xi$s are strong-interaction and weak phases, respectively.
They contribute to the $CP$ asymmetry~\cite{Donoghue:1986hh}
\begin{align} \label{acp}
A_{CP}^{} & \,=\, \frac{\alpha+\overline\alpha}{\alpha-\overline\alpha} \,\simeq\, -\tan \big( \delta_{1\mathbb B}
- \delta_{1\mathbb A} \big) \tan\big(\xi_{1\mathbb B}^{}-\xi_{1\mathbb A}^{}\big) \,, &
\end{align}
where \,$\alpha=2\tilde\kappa\,{\rm Re}\big(\mathbb A^*\,\mathbb B\big)/
\big(|\mathbb A|^2 + |\tilde\kappa\,\mathbb B|^2\big)$\,
and $\overline\alpha$ stands for its antiparticle counterpart, with the kinematical factor \,$\tilde\kappa=[(E_{\rm f}-m_{\rm f})/(E_{\rm f}+m_{\rm f})]^{1/2}$\, depending on the energy $E_{\rm f}$ and mass $m_{\rm f}$ of the emitted baryon in the rest frame of the parent hyperon.

Numerically, with the central values of the various input parameters, from eqs.\,\,(\ref{Aamplitudes}) and (\ref{Bamplitudes}) we find the \,$\Delta I=1/2$\, components
\begin{align}
\mathbb A_{1/2}^{\Lambda\rm,new} & \,=\,
0.134\, {\cal C}_-^{} + 0.135\, {\cal C}_+^{} - 0.244\, {\cal C}_u^{} \,, &
\mathbb B_{1/2}^{\Lambda\rm,new} & \,=\, 0.980\, {\cal C}_-^{} + 0.981\, {\cal C}_+^{} + 2.15\, {\cal C}_u^{} \,,
\nonumber \\
\mathbb A_{1/2}^{\Xi\rm,new} & \,=\, 0.146\, {\cal C}_-^{} + 0.146\, {\cal C}_+^{} - 0.300\, {\cal C}_u^{} \,, &
\mathbb B_{1/2}^{\Xi\rm,new} & \,=\, 0.349\, {\cal C}_-^{} + 0.469\, {\cal C}_+^{} + 0.680\, {\cal C}_u^{} \,,
\end{align}
all the numbers in $\rm GeV^2$.
The corresponding quantities inferred from hyperon data \cite{ParticleDataGroup:2024cfk} are
\begin{align} \label{expAB}
\mathbb A_{1/2}^{\Lambda\rm,exp} & \,=\, -0.998\pm0.005 \,, & \mathbb B_{1/2}^{\Lambda\rm,exp} & \,=\, -7.96\pm0.06 \,, &
\nonumber \\
\mathbb A_{1/2}^{\Xi\rm,exp} & \,=\, -1.447\pm0.007 \,, & \mathbb B_{1/2}^{\Xi\rm,exp} & \,=\, 4.46\pm0.06 \,,
\end{align}
all in units of $G_{\rm F}^{}m_{\pi^+}^2$.
Following the usual practice~\cite{Donoghue:1986hh,Chang:1994wk,Deshpande:1994vp,He:1995na,He:1999bv,Tandean:2002vy,Tandean:2003fr}, we approximate each weak phase by taking the ratio of the theoretical amplitude to its experimental counterpart.
Thus, with the CKM matrix elements \,$V_{ud}=0.97435(16)$\, and \,$V_{us}=0.22501(68)$  \cite{ParticleDataGroup:2024cfk},
as well as the empirical strong phase shifts \,$\delta_{1\mathbb B}^\Lambda-\delta_{1\mathbb A}^\Lambda=-7.3^\circ\pm0.1^\circ$ \cite{Salone:2022lpt,Hoferichter:2015hva}
and \,$\delta_{1\mathbb B}^\Xi-\delta_{1\mathbb A}^\Xi = 1.7^\circ\pm1.1^\circ$ \cite{ParticleDataGroup:2024cfk}, from the central values we obtain the new contributions to the weak-phase differences
\begin{align} \label{xinp}
\xi_{1\mathbb B}^{\Lambda\rm,new}
{-\,} \xi_{1\mathbb A}^{\Lambda\rm,new}
& = {\rm Im} \Bigg[ \frac{\mathbb B_{\scriptscriptstyle1/2}^{\Lambda\rm,new}}
{\mathbb B_{\scriptscriptstyle1/2}^{\Lambda\rm,exp}} - \frac{\mathbb A_{\scriptscriptstyle1/2}^{\Lambda\rm,new}}
{\mathbb A_{\scriptscriptstyle1/2}^{\Lambda\rm,exp}} \Bigg]
= \frac{{\rm Im}\big[ \big( 0.053 \mathbb Y_{dd}^{} - 0.113 \mathbb Y_{ss}^{} \big) \mathbb Y_{sd}^* - 0.497 \mathbb Y_{dd}^* \mathbb Y_{ss}^{} \big]}{m_H^2/\rm TeV^2} \,,
\nonumber \\
\xi_{1\mathbb B}^{\Xi\rm,new} {-\,} \xi_{1\mathbb A}^{\Xi\rm,new} & = {\rm Im} \Bigg[ \frac{\mathbb B_{\scriptscriptstyle1/2}^{\Xi\rm,new}}
{\mathbb B_{\scriptscriptstyle1/2}^{\Xi\rm,exp}}
- \frac{\mathbb A_{\scriptscriptstyle1/2}^{\Xi\rm,new}}
{\mathbb A_{\scriptscriptstyle1/2}^{\Xi\rm,exp}} \Bigg]
= \frac{{\rm Im}\big[ \big( 0.847 \mathbb Y_{dd}^{} + 0.047 \mathbb Y_{ss}^{} \big) \mathbb Y_{sd}^* - 0.053 \mathbb Y_{dd}^* \mathbb Y_{ss}^{} \big]}{m_H^2/\rm TeV^2}\,,
\end{align}
and hence to the $A_{CP}$ asymmetries
\begin{align} \label{ACPnp}
A_{CP}^{\Lambda\rm,new} & \,=\, \Big( \xi_{1\mathbb B}^{\Lambda\rm,new} - \xi_{1\mathbb A}^{\Lambda\rm,new} \Big) \tan7.3^\circ \,=\, 10^3\, {\rm Im} \big( 7\,{\cal C}_-^{} + 7\, {\cal C}_+ - 290\, {\cal C}_u \big) \rm GeV^2 \,,
\nonumber \\
A_{CP}^{\Xi\rm,new} & \,=\, -\Big( \xi_{1\mathbb B}^{\Xi\rm,new} - \xi_{1\mathbb A}^{\Xi\rm,new} \Big) \tan1.7^\circ \,=\, 10^3\, {\rm Im} \big( {-}23.4\,{\cal C}_-^{} - 26.9\, {\cal C}_+ + 7.1\, {\cal C}_u \big) \rm GeV^2 \,. ~~~
\end{align}
It is worth remarking that these calculations involve significant uncertainties, related to those in the theoretical treatment of hyperon nonleptonic decays~\cite{Donoghue:1985rk,Bijnens:1985kj,Tandean:2002vy,Tandean:2003fr}.
For comparison, the SM predicts \cite{Tandean:2002vy,Salone:2022lpt}
\begin{align}
-2.4 & \,\le\, 10^4 \big( \xi_{1\mathbb B}^{\Lambda}
- \xi_{1\mathbb A}^{\Lambda} \big)_{\textsc{sm}} \,\le\, +2.0 \,, &
-3.8 & \,\le\, 10^4 \big( \xi_{1\mathbb B}^{\Xi}
- \xi_{1\mathbb A}^{\Xi} \big)_{\textsc{sm}} \,\le\, -0.3 \,,
\nonumber \\
-3 & \,\le\, 10^5
A_{CP}^{\Lambda\rm,\textsc{sm}} \,\le\, 3 \,, & 0 & \,\le\, 10^5
A_{CP}^{\Xi\rm,\textsc{sm}} \,\le\, 2 \,, \label{ACPsm}
\end{align}
the $A_{CP}^{\Xi\rm,\textsc{sm}}$ result having been updated with the $\delta_{1\mathbb B}^\Xi{-}\delta_{1\mathbb A}^\Xi$ value quoted above,
and the pertinent $A_{CP}$ data are \cite{ParticleDataGroup:2024cfk}
\begin{align} &
A_{CP}^{\Lambda\rm,exp} \,=\, (-1\pm4)\times10^{-3} \,, & &  A_{CP}^{\Xi^-\rm,exp}  \,=\, (6\pm14)\times10^{-3} ~~ \mbox{\cite{BESIII:2021ypr}} \,, &
\nonumber \\ &
A_{CP}^{\Lambda\rm,exp}+A_{CP}^{\Xi^-\rm,exp} \,=\, (0\pm7)\times10^{-4} ~~ \mbox{\cite{HyperCP:2004zvh}} \,, \label{ACPexp}
\end{align}
which are consistent with zero.\footnote{Pursuits of $A_{CP}^{\Lambda}$, $A_{CP}^{\Xi}$, and their sum have been performed in many experiments~\cite{R608:1985fmh,Barnes:1987vc,DM2:1988ppi,Barnes:1996si,E756:2000rge,HyperCP:2004zvh,BESIII:2018cnd,BESIII:2022qax,BESIII:2021ypr,BESIII:2022lsz,BESIII:2022lsz,BESIII:2023lkg,BESIII:2023drj,BESIII:2023jhj,BESIII:2024nif,Belle:2022uod,LHCb:2024tnq}, including most recently by BESIII~\cite{BESIII:2018cnd,BESIII:2021ypr,BESIII:2022qax,BESIII:2022lsz,BESIII:2023lkg,BESIII:2023drj,BESIII:2023jhj,BESIII:2024nif}, Belle~\cite{Belle:2022uod}, and LHCb~\cite{LHCb:2024tnq}.
Searches for $CP$ violation in $\Sigma$ hyperon decays has also been carried out by BESIII~\cite{BESIII:2024nif,BESIII:2020fqg,BESIII:2023sgt}.\bigskip}

\subsection{Kaon processes\label{kaons}}

\begin{figure}[t] \centering
\includegraphics[trim=43mm 246mm 43mm 45mm,clip,width=0.97\textwidth]{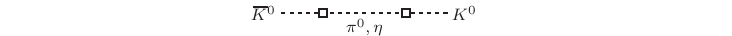} \vspace{-1ex}
\caption{Diagrams for long-distance contribution to kaon mixing with either both hollow squares or one of them symbolizing couplings from ${\cal L}_\chi^{\rm new}$ in eq.\,(\ref{Lchi}), in the latter case the other square representing the SM weak coupling from ${\cal L}_{\rm w}^{\textsc{sm}}$.} \label{Kmixing}
\end{figure}

The same Yukawa couplings, $\mathbb Y_{dd,sd,ss}$, can  influence neutral-kaon mixing at long distance (LD) via the diagrams exhibited in figure\,\,\ref{Kmixing} mediated by $\pi^0$ and \,$\eta=\eta_8^{}$,\, where either both vertices come from ${\cal L}_\chi^{\rm new}$ or only one of them does and the other one is furnished by the SM weak chiral Lagrangian \,${\cal L}_{\rm w}^{\textsc{sm}}\supset\gamma_8^{}f_\pi^2\, {\rm Tr}\big( \lambda_6^{}\, \partial_\nu^{} \Sigma\, \partial^\nu \Sigma^\dagger \big)$,  where \,$\gamma_8^{}=7.7\times 10^{-8}$\, from \,$K\to\pi\pi$\, data~\cite{Cirigliano:2011ny} and $\lambda_6$ is the sixth Gell-Mann matrix acting in flavor space.\footnote{Heavier unflavored mesons, such as the $\eta'$, can also contribute, but their effects are expected to be relatively smaller due to their bigger masses in the propagators.}
The resulting LD contribution to the amplitude for $~\overline{\!K}{}^0\to K^0$\, is
\begin{align}
{\cal M}_{\bar K\to K}^{\rm new}\, & =\, \frac{ 4\gamma_8^{} f_\pi^2\, \tilde{\textsc c}{}_-^{}\, \hat{\textsc h}_-^{}\, m_K^2\, \big(m_\pi^2-m_\eta^2\big) + 2\, \tilde{\textsc c}{}_-^2\, \hat{\textsc h}{}_-^2\, \big( 4m_K^2-3m_\pi^2-m_\eta^2\big) }{f_\pi^4\, \big(m_K^2-m_\pi^2\big)\big(m_K^2-m_\eta^2\big)} \,.
\end{align}
This modifies the neutral-kaon mass difference \,$\Delta M_K^{} = {\rm Re}\,{\cal M}_{\bar K\to K}/m_K^{}$\, and the indirect kaon $CP$-violation parameter \,$|\varepsilon| = \big| {\rm Im}\,{\cal M}_{\bar K\to K} \big|/\big(\sqrt8\, m_K^{}\, \Delta M_K^{\rm exp}\big)$.\,
Numerically we arrive at
\begin{align} \label{DMKnew}
\Delta M_K^{\rm new}\, & =\, {\rm Re} \big( 1.29\times10^{-7}\, {\cal C}_-^{}{\rm~GeV}^3 - 0.283\, {\cal C}_-^2{\rm~GeV}^5 \big)
\nonumber \\ & =\, 10^{-14}\, {\rm Re} \Bigg\{ \frac{6.47\,  \mathbb Y_{sd}^*
\big(\mathbb Y_{dd}^{}-\mathbb Y_{ss}^{}\big) }{(m_H/\rm TeV)^2} - \frac{7.07\,
\big[ \mathbb Y_{sd}^* \big(\mathbb Y_{dd}^{}-\mathbb Y_{ss}^{}\big) \big]\raisebox{1pt}{$^2$} }{(m_H/\rm TeV)^4} \Bigg\} \rm GeV \,, ~~~
\\ \label{epsnew} \raisebox{4ex}{}
 \varepsilon_{\rm new}\, & =\, {\rm Im} \big( 1.31\times10^7\, {\cal C}_-^{}{\rm~GeV}^2 - 2.87\times10^{13}\, {\cal C}_-^2{\rm~GeV}^4 \big)
\nonumber \\ & =\, \frac{ 6.56~ {\rm Im} \big[ \mathbb Y_{sd}^* \big( \mathbb Y_{dd}^{} - \mathbb Y_{ss}^{} \big) \big] }{(m_H/\rm TeV)^2} - \frac{ 7.18~ {\rm Im} \big[ \mathbb Y_{sd}^*
\big( \mathbb Y_{dd}^{} - \mathbb Y_{ss}^{} \big) \big]{}^2 }{(m_H/\rm TeV)^4} \,.
\end{align}

The new interactions in eq.\,(\ref{Lchi}) also affect direct $CP$ violation in \,$K\to\pi\pi$.\,
This is represented by the contact and tadpole diagrams in figure \ref{diagrams}(c), giving rise to the amplitudes
\begin{align} \label{K2pipi}
{\cal M}_{K^0\to\pi^+\pi^-}^{\rm new} & =
\frac{i\sqrt2}{f_\pi^3} \big( {-}\tilde{\textsc c}{}_u^* \hat{\textsc h}_u^{} - \tilde{\textsc c}{}_+^* \hat{\textsc h}_+^{} - \tilde{\textsc c}{}_-^* \hat{\textsc h}_-^{} \big)
= \sqrt2\, {\cal M}_{K^+\to\pi^-\pi^0}^{\rm new} \,, &
{\cal M}_{K^0\to\pi^0\pi^0}^{\rm new} & = 0 \,.
\end{align}
In terms of their \,$\Delta I=1/2$ and 3/2\, components
\,$\widetilde{\texttt M}_{1/2}$ and
\,$\widetilde{\texttt M}_{3/2}$, the \,$\Delta I=5/2$\, one being neglected, the amplitudes are generally expressible as~\cite{Cirigliano:2011ny}
\begin{align}
-i{\cal M}_{K^0\to\pi^+\pi^-}^{} & \,=\,
\widetilde{\texttt M}_{1/2}^{} + \tfrac{1}{\sqrt2}\,
\widetilde{\texttt M}_{3/2}^{} \,, &
-i{\cal M}_{K^0\to\pi^0\pi^0}^{} & \,=\,
\widetilde{\texttt M}_{1/2}^{} - \sqrt2~
\widetilde{\texttt M}_{3/2}^{} \,, &
\nonumber \\
-i{\cal M}_{K^+\to\pi^-\pi^0}^{} & \,=\, \tfrac{3}{2}\,
\widetilde{\texttt M}{}_{3/2}^{} \,.
\end{align}
Thus, from eq.\,(\ref{K2pipi})
\begin{align}
\widetilde{\texttt M}{}_{1/2}^{\rm new} & \,=\,
\sqrt2~ \widetilde{\texttt M}{}_{3/2}^{\rm new} \,=\, \frac{2\sqrt2}{3f_\pi^3} \big( {-}\tilde{\textsc c}{}_u^* \hat{\textsc h}_u^{} - \tilde{\textsc c}{}_+^* \hat{\textsc h}_+^{} - \tilde{\textsc c}{}_-^* \hat{\textsc h}_-^{} \big) \,. &
\end{align}
The contribution of
\,$\widetilde{\texttt M}{}_{1/2,3/2}^{\rm new}$ to $\varepsilon'$ can be expressed as
\begin{align} \label{e'new}
\frac{\varepsilon_{\rm new}'}{\varepsilon} & \,=\, \frac{\omega}{\sqrt2\, \big|\varepsilon_{\rm exp}\big|} \Bigg( \frac{{\rm Im} A_2^{\rm new}}{{\rm Re}\, A_2^{\rm exp}} - \frac{{\rm Im} A_0^{\rm new}}{{\rm Re}\, A_0^{\rm exp}} \Bigg) \,, &
\end{align}
with the empirical values \cite{Cirigliano:2011ny}
\begin{align} \label{Aexp}
{\rm Re}\, A_0^{\rm exp} & \,=\, (2.704\pm0.001) \times10^{-7} \rm\,GeV \,, & {\rm Re}\, A_2^{\rm exp} & \,=\, (1.210 \pm 0.002) \times10^{-8} \rm\,GeV \,,
\end{align}
where \,$\omega = {\rm Re}\,A_2^{\rm exp}/{\rm Re}\,A_0^{\rm exp} \simeq 1/22$\,  and \,$A_{0,2}^{\rm new} = \widetilde{\texttt M}{}_{1/2,3/2}^{\rm new}$,\, leading to
\begin{align}
\frac{\varepsilon_{\rm new}'}{\varepsilon}\, & =\, \frac{1-\sqrt2\,\omega}{\sqrt2\, \big|\varepsilon_{\rm exp}\big|}~ \frac{\rm Im\,
\widetilde{\texttt M}{}_{3/2}^{new}}{{\rm Re}\, A_0^{\rm exp}} \,=\, {\rm Im} \big( 2.97\, {\cal C}_-^{} + 2.97\, {\cal C}_+^{} + 2.84\, {\cal C}_u^{} \big) \times10^8\rm~GeV^2
\nonumber \\ & =\, \frac{{\rm Im} \big[ \big( 297\, \mathbb Y_{dd}^{} + 14\, \mathbb Y_{ss}^{} \big) \mathbb Y_{sd}^* + 62\, \mathbb Y_{dd}^* \mathbb Y_{ss}^{} \big) \big]}{m_H^2/\rm TeV^2} \,. & \label{eps'new}
\end{align}

\subsection{Parameter scans\label{scans}}

\begin{figure}[b!] \bigskip \centering
\includegraphics[width=0.4\linewidth]{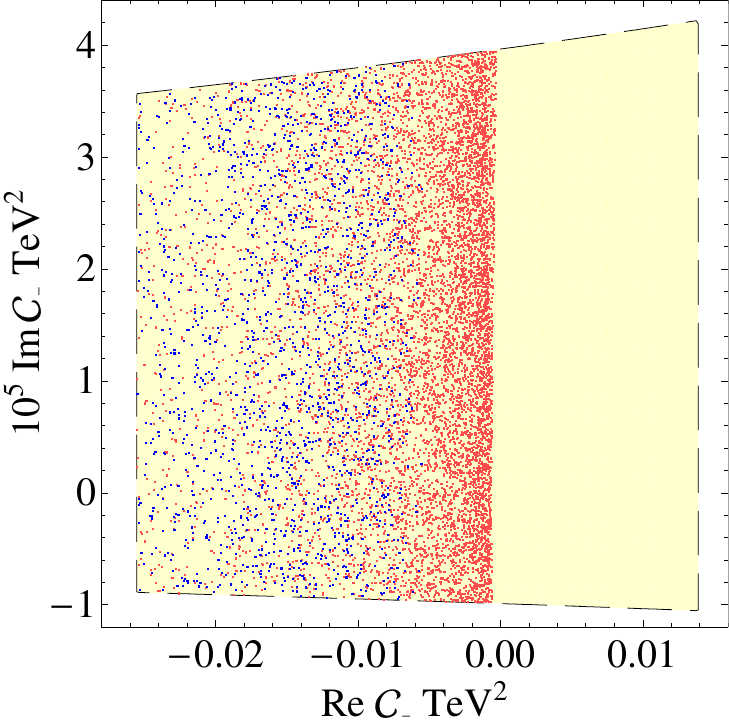} ~ ~ ~
\includegraphics[width=0.4\linewidth]{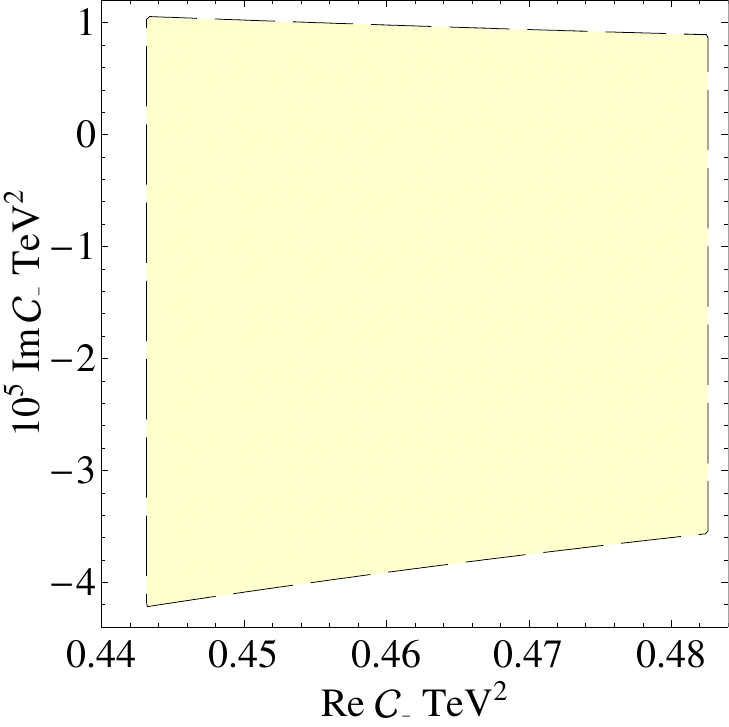}
\caption{Parameter space in Re$\,{\cal C}_-$-Im$\,{\cal C}_-$ plane (lightly shaded areas bounded by dashed curves on both panels) allowed by $\Delta M_K$ and $\varepsilon$ data.
The red and blue points, appearing only on the left plot, correspond to the new contributions to the $CP$ asymmetry $A_{CP}^\Lambda$ falling within the ranges \,$10^{-4} \le \big|A_{CP}^{\Lambda\rm,new}\big| < 10^{-3}$ and \,$10^{-3} \le \big|A_{CP}^{\Lambda\rm,new}\big| \le 3.5 \times10^{-3}$, respectively, and fulfill all the dark-matter and kaon requirements.} \label{fig:mixcon}
\end{figure}

Before turning to the values of $\mathbb Y_{\textit{\texttt{qq}}'}$, \,$\textit{\texttt{qq}}'=dd,sd,ss$,\, meeting various requisites, it is instructive to look at the complementarity of the kaon and hyperon quantities treated above for probing the $CP$ violation due to $\,{\cal C}_{\mp,u}$.
We start with \,$\Delta M_K^{\rm new} = 10^{-15}\, {\rm Re} \big( 129\, {\cal C}_-^{}{\rm TeV}^2 - 283\, {\cal C}_-^2{\rm TeV}^4 \big)$GeV\, from eq.\,(\ref{DMKnew}) and \,$\varepsilon_{\rm new} = {\rm Im} \big( 13.1\, {\cal C}_-^{}{\rm TeV}^2 - 28.7\, {\cal C}_-^2 {\rm TeV}^4 \big)$\, from eq.\,(\ref{epsnew}).
Then, in view of the measured \,$\Delta M_K^{\rm exp} = (3.484\pm0.006)\times10^{-15}$\,GeV \cite{ParticleDataGroup:2024cfk} and SM estimate \,$\Delta M_K^{\textsc{sm}} = (5.8\pm2.4)\times10^{-15}$\,GeV\, from lattice QCD work~\cite{Wang:2022lfq}, we may require \,$-1<\Delta M_K^{\rm new}/\Delta M_K^{\rm exp}<0.5$,\, which is in line with the 2$\sigma$ range of \,$\Delta M_K^{\rm exp}$\,$-$\,$\Delta M_K^{\textsc{sm}}$.\,
Moreover, comparing the data  \,$\big|\varepsilon_{\rm exp}\big|=(2.228\pm0.011)\times10^{-3}$~\cite{ParticleDataGroup:2024cfk} and \,$|\varepsilon_{\textsc{sm}}|=(2.030\pm0.162)\times10^{-3}$\,  calculated on the lattice~\cite{Jwa:2023uuz}, from the 2$\sigma$ range of their difference we can impose \,$-1.3 < 10^4\, \varepsilon_{\rm new} < 5.2$.\,
Putting these things together, we obtain the lightly shaded areas bounded by dashed curves on the two plots of Im$\,{\cal C}_-$ versus Re$\,{\cal C}_-$ in figure\,\,\ref{fig:mixcon} and, as evident therein, \,$|{\rm Im}\cal\,C_-\rm| < 4.2 \times10^{-5}\rm\,TeV^{-2}$.\,

Next, from eq.\,(\ref{eps'new}) we have \,$\varepsilon_{\rm new}'/\varepsilon = 297\, {\rm Im}
\big( {\cal C}_-^{} + {\cal C}_{+u} \big)\rm TeV^2$,\, where \,${\cal C}_{+u} \equiv {\cal C}_+ + 0.959\,{\cal C}_u$.\,
It is bounded by \,$-2.2 < 10^3\,   \varepsilon_{\rm new}'/\varepsilon < 1.2$,\, inferred from \,${\rm Re}(\varepsilon'/\varepsilon)_{\rm exp}=(16.6\pm2.3)\times10^{-4}$ \cite{ParticleDataGroup:2024cfk} and  \,${\rm Re}(\varepsilon'/\varepsilon)_{\textsc{sm}}=(21.7\pm8.4)\times10^{-4}$  \cite{RBC:2020kdj} at the 2$\sigma$ level.
With the constraints on ${\cal C}_-^{}$ as seen in figure\,\,\ref{fig:mixcon} taken into account, it is then straightforward to arrive at \,$-5.0 < 10^5\,{\rm Im}\,{\cal C}_{+u}\rm\,TeV^2<4.6$.\,

For the hyperons, from eq.\,(\ref{ACPnp}) we can write \,$A_{CP}^{\Lambda\rm,new} =  10^{-3}\, {\rm Im} \big( 7\,{\cal C}_- - 141\, {\cal C}_{+u} + 148\, {\cal C}_{-u} \big) \rm TeV^2$\, and \,$A_{CP}^{\Xi\rm,new} = 10^{-3}\, {\rm Im} \big( {-}23.4\,{\cal C}_- - 10.5\, {\cal C}_{+u} - 16.5\, {\cal C}_{-u} \big) \rm TeV^2$,\, where  \,${\cal C}_{-u} \equiv {\cal C}_+ - {\cal C}_u/0.959$.\,
Accordingly, as we just learned that \,$|{\rm Im}\,{\cal C}_{-,+u}|\mbox{\footnotesize~$\lesssim$~}5\times10^{-5}\rm\,TeV^{-2}$,\,
for $\,{\cal C}_{-u}$ vanishing \,$\big|A_{CP}^{\Lambda\rm,new}\big| < 8\times10^{-6}$\, and \,$\big|A_{CP}^{\Xi\rm,new}\big| < 2\times10^{-6}$,\, whereas for Im$\,{\cal C}_{-u}$ being nonzero and much larger than $10^{-5}$ we have approximately \,$A_{CP}^{\Lambda\rm,new} = 0.148\, {\cal C}_{-u}\rm\,TeV^2$\, and \,$A_{CP}^{\Xi\rm,new} = -0.0165\, {\cal C}_{-u}{\rm\,TeV^2} = -0.111\,A_{CP}^{\Lambda\rm,new}$.\,
Thus, $A_{CP}$ can be sensitive to a direction in parameter space unconstrained by the kaon measurements.
Furthermore, if \,${\cal C}_{-u}{\rm\,TeV^2}$ is capped at unity to ensure perturbativity, $A_{CP}^{\Lambda\rm,new}$ can potentially become huge, up to around 15\%.
However, as we address next, the underlying model parameters that make up \,${\cal C}_{-u}$ must comply with other restrictions, rendering its size smaller.

We see in eq.\,(\ref{C-+u}) that ${\cal C}_{-,+,u}$ depend on the Yukawa couplings $\mathbb Y_{\textit{\texttt{qq}}'}$ and the heavy Higgs mass~$m_H$.
Since these parameters determine the coefficients $C_{dD}^{S,\textit{\texttt{qq}}'}$ in eq.\,(\ref{CdDS}) as well, which characterize  the darkon-quark effective interactions, restrictions from DM relic density and direct-search data are also relevant, as detailed in section \ref{DMsector}.
To illustrate the $\mathbb Y_{\textit{\texttt{qq}}'}$ values that can translate into $A_{CP}$ exceeding its maximal SM expectations of order $10^{-5}$ as quoted in eq.\,(\ref{ACPsm}), we pick for definiteness a representative point from figure \ref{fig:dm} specified by
\begin{align} \label{dm_reqs}
R_{d/s} & \,=\, \frac{{\rm Re}\, \mathbb Y_{dd}}{{\rm Re}\, \mathbb Y_{ss}} \,=\, -0.04 \,, & \big|\hat R_{d/s}\big| & \,=\, \frac{|\mathbb Y_{sd}|}{2\,|{\rm Re}\, \mathbb Y_{ss}|} \,=\, 1 \,, &
\big|C_{dD}^{S,ss}\big| & \,=\, \frac{0.08}{\rm TeV} \,,
\end{align}
and \,$m_H=1$\,TeV.\,
Another requisite is that the Yukawa and darkon-$H$ interactions remain  perturbative, and so we demand \,$\big|\mathbb Y_{\textit{\texttt{qq}}'}\big|<\sqrt{4\pi}$\, and \,$|\lambda_3|<4\pi$.\,
Lastly, the kaon decay amplitudes provide additional important bounds.
Particularly, comparing  \,${\rm Re}\, A_0^{\textsc{sm}} = (2.44\pm0.55) \times10^{-7}$\,GeV  and \,${\rm Re}\, A_2^{\textsc{sm}} = (1.23 \pm 0.41) \times10^{-8}$\,GeV  \cite{RBC:2020kdj}  to their empirical counterparts in eq.\,(\ref{Aexp}), we can derive \,$-0.8 < 10^7\,{\rm Re}\, A_0^{\rm new}\rm/GeV < 1.3$\, and \,$ 10^9\,\big|{\rm Re}\, A_2^{\rm new}\big| < 8.0$\,\,GeV.\,
The corresponding contributions to $\mathbb A$ and $\mathbb B$ in the hyperon amplitudes described in section \ref{hyperons} are also subject to restrictions, but, since their computation is still far from precise~\cite{Donoghue:1985rk,Bijnens:1985kj,Tandean:2002vy,Tandean:2003fr}, the preceding Re\,$A_{0,2}$ constraints, especially the latter, are more consequential.

\begin{figure}[b!] \bigskip \centering
\includegraphics[width=0.4\linewidth]{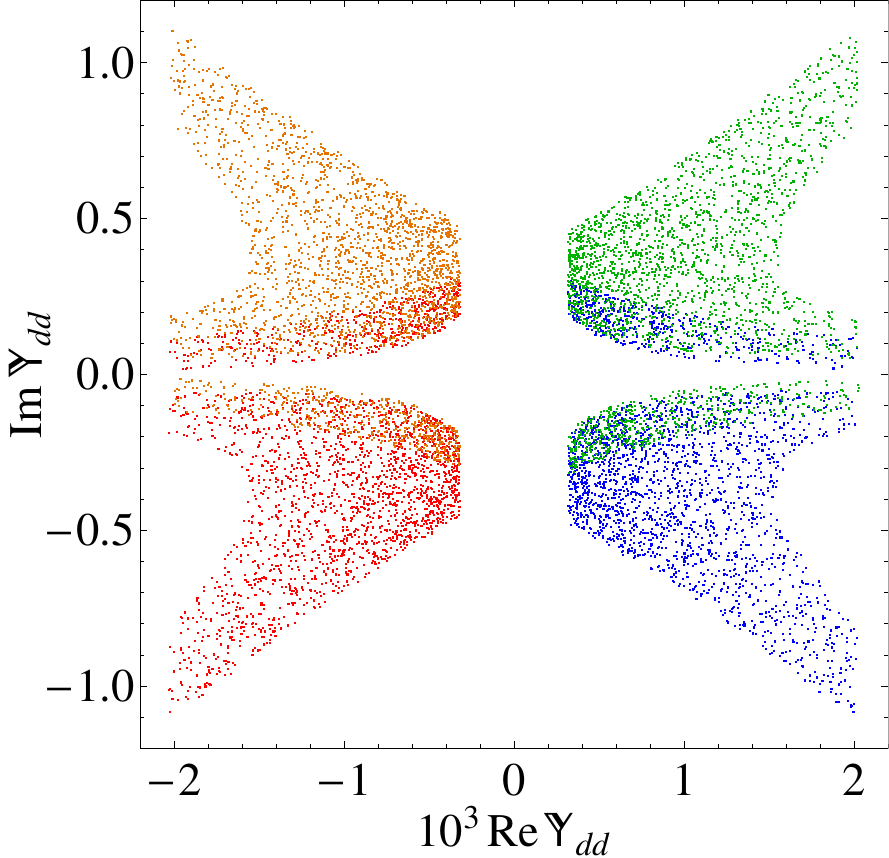} ~ ~ ~ ~
\includegraphics[width=0.4\linewidth]{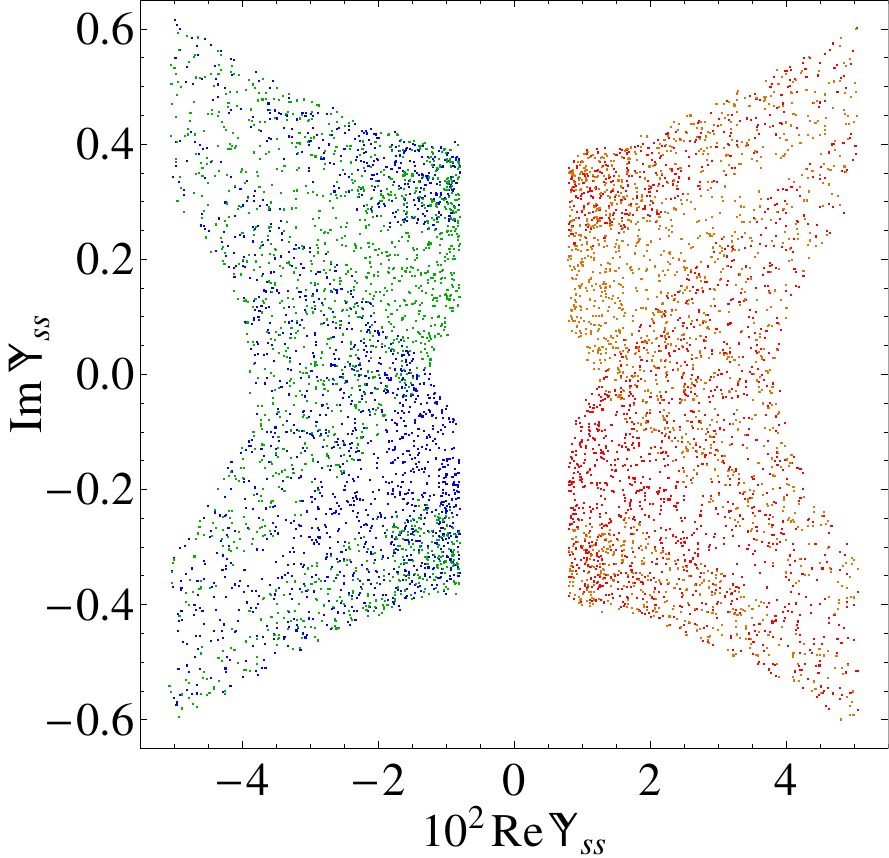}
\caption{The imaginary parts of $\mathbb Y_{dd}$ (left) and $\mathbb Y_{ss}$ (right) versus their respective real parts, all of which comply with the dark-matter and kaon requirements.
The colored points in this and the next two figures correspond to \,${\rm arg}\,\mathbb Y_{sd} \simeq 84^\circ$ (red), $-84^\circ$ (orange), $96^\circ$ (blue), and $-96^\circ$ (green).} \label{fig:imY}
\end{figure}

After generating numerous sets of random numbers for $\mathbb Y_{\textit{\texttt{qq}}'}$ which fulfill all of the requirements described above, we display in
figure \ref{fig:imY} the outcomes which can translate into \,$\big|A_{CP}^{\Lambda\rm,new}\big|\ge10^{-4}$.\,
This is achievable with \,$|{\rm arg}\,\mathbb Y_{sd}|\simeq84^\circ$\, or \,$\pi-84^\circ$,\, as indicated in the figure, and within very narrow ranges around these phase values, beyond which $\big|A_{CP}^{\Lambda\rm,new}\big|$ quickly becomes much smaller.
Thus, the sizable asymmetries can be reached with a high degree of fine tuning.

\begin{figure}[!b] \centering
\includegraphics[width=0.358\linewidth]{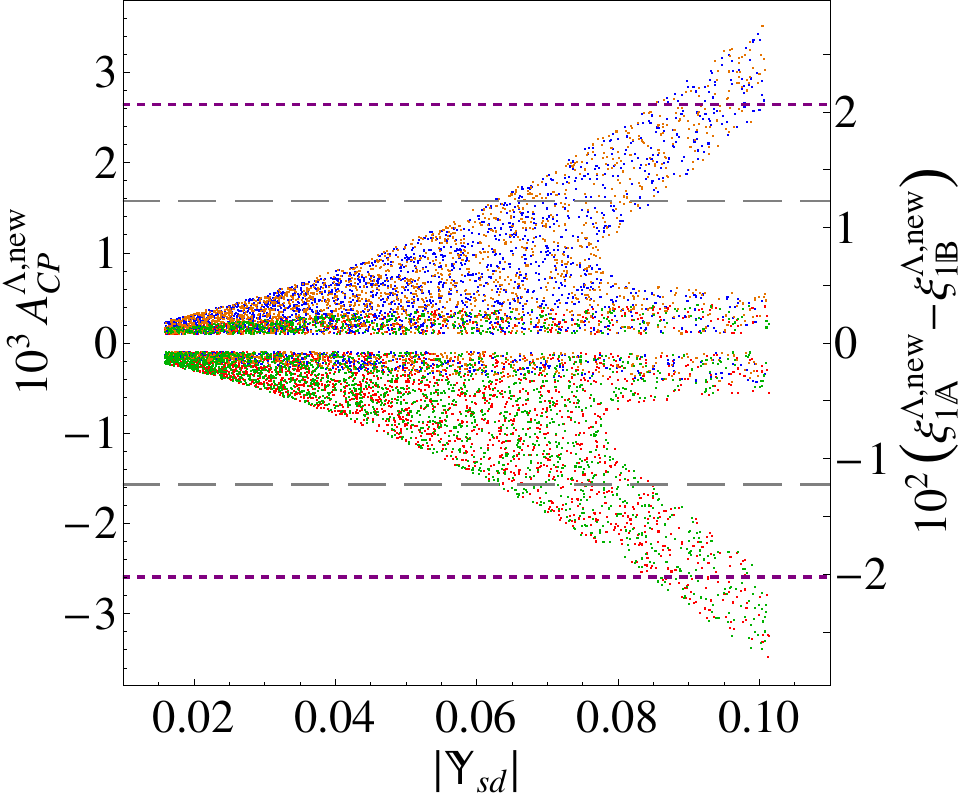} ~ \includegraphics[width=0.307\linewidth]{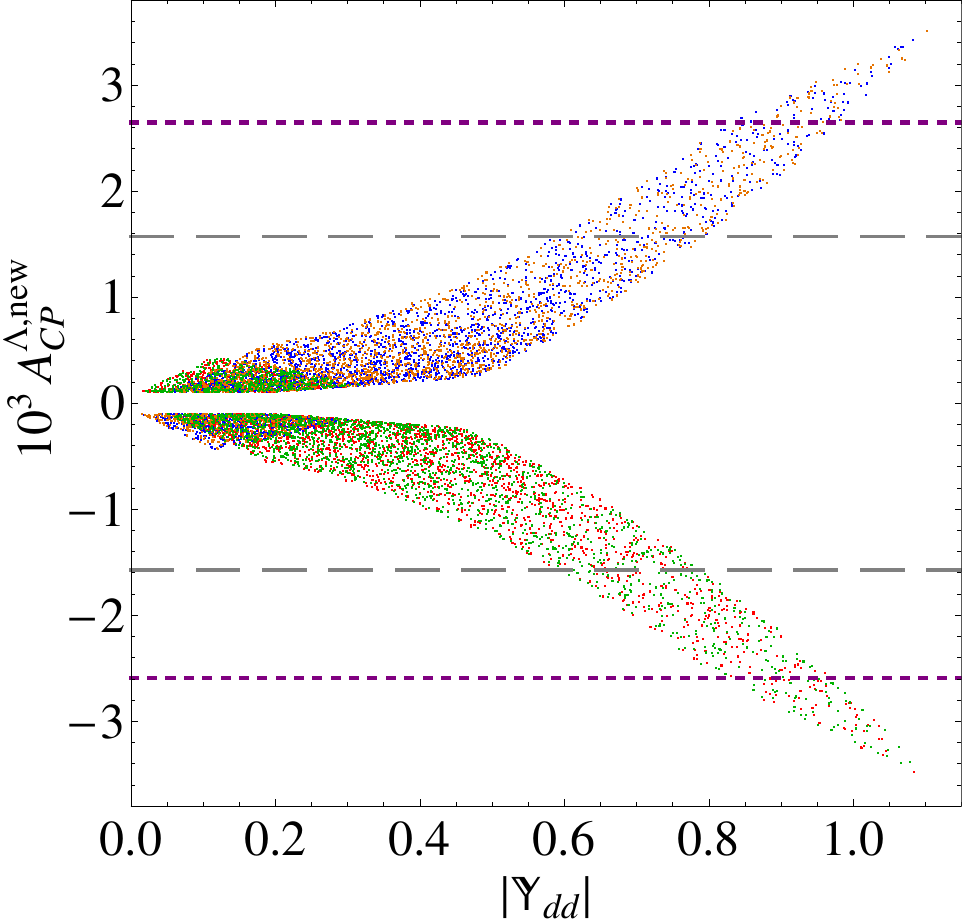}~\includegraphics[width=0.305\linewidth]{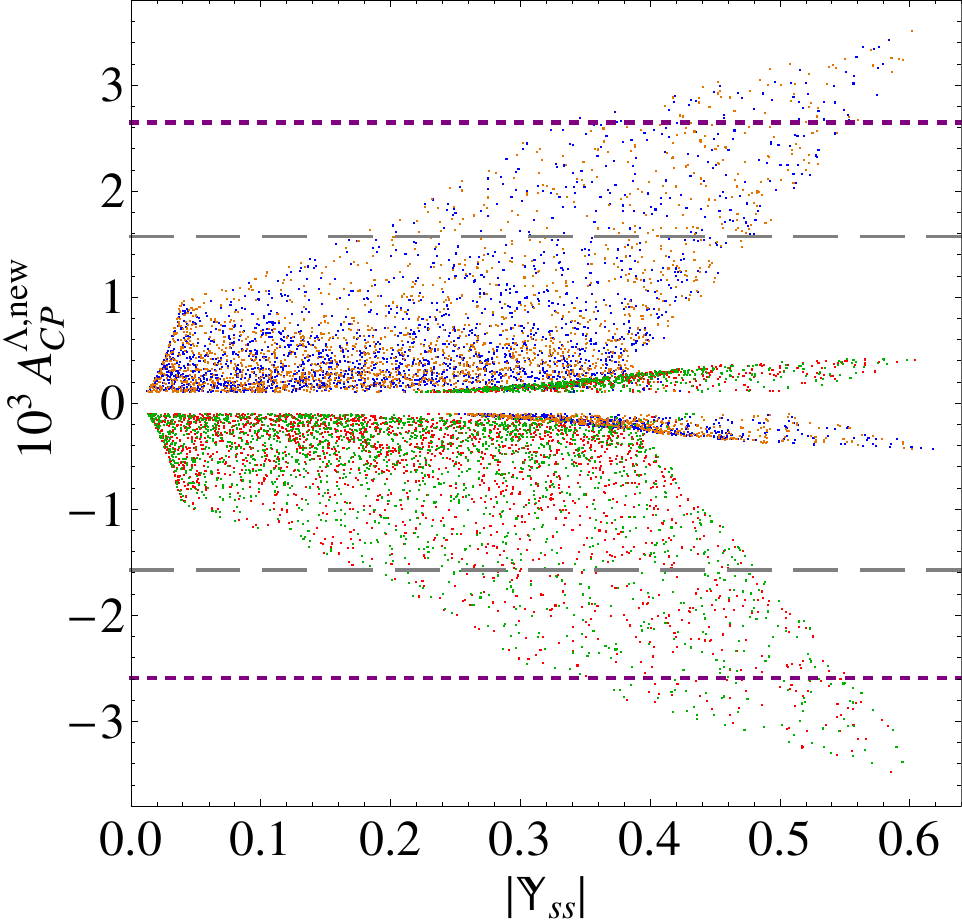}\vspace{2ex}\\
\includegraphics[width=0.4\linewidth]{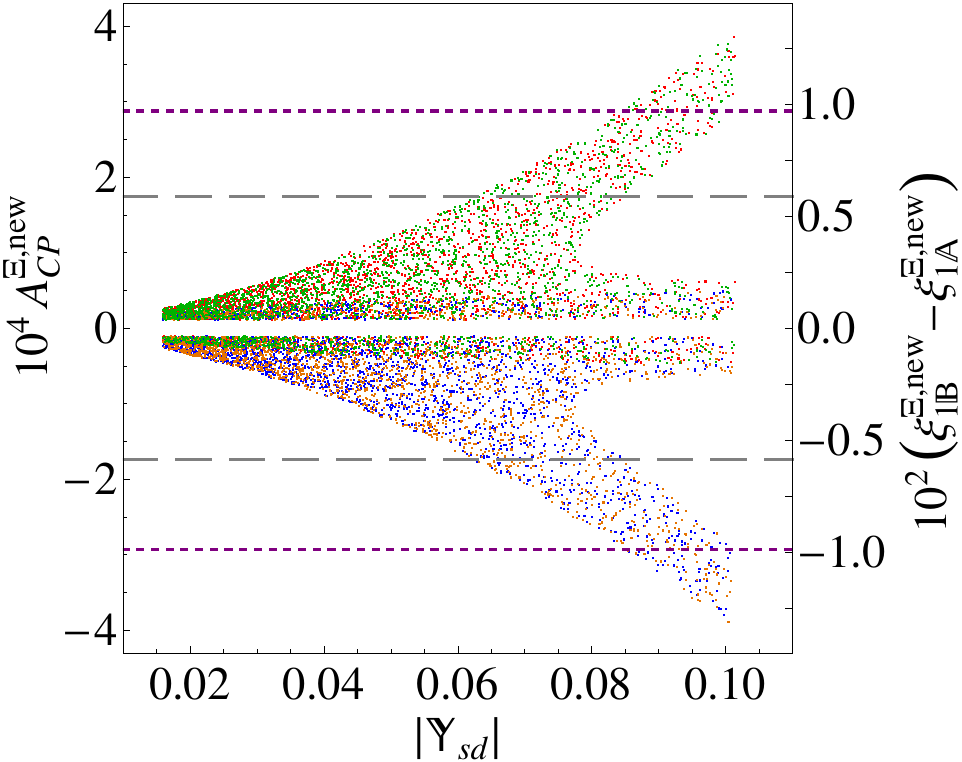} ~ ~ ~ ~ \includegraphics[width=0.33\linewidth]{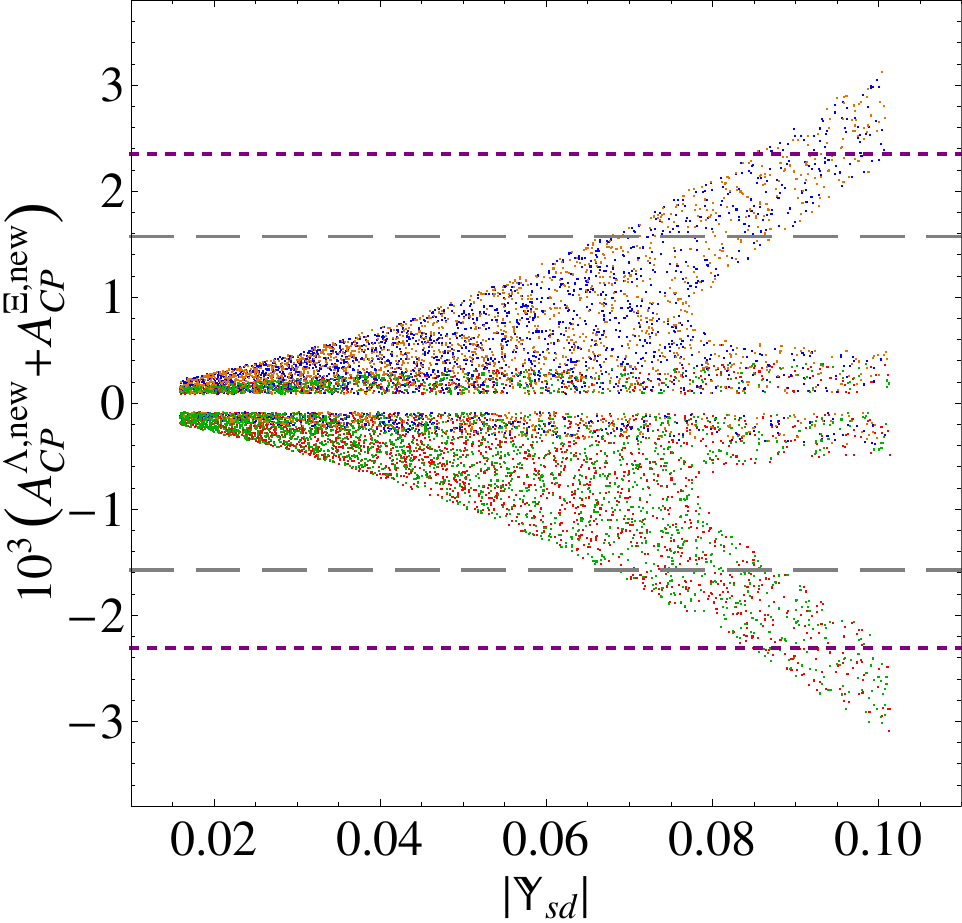}\vspace{-2pt}
\caption{Top: the ranges of $A_{CP}^{\Lambda\rm\,new}$ in relation to the values of $|\mathbb Y_{sd}|$, $|\mathbb Y_{dd}|$, and $|\mathbb Y_{ss}|$ satisfying the DM and kaon constraints.
Bottom: the corresponding $A_{CP}^{\Xi\rm\,new}$ and \,$A_{CP}^{\Lambda\rm,new}{+}A_{CP}^{\Xi\rm,new}$  versus $|\mathbb Y_{sd}|$.
The top- and bottom-left panels also show the weak-phase differences \,$\xi_{1\mathbb A}^{\Lambda\rm,new}{-}\,\xi_{1\mathbb B}^{\Lambda\rm,new}$\, and
\,$\xi_{1\mathbb B}^{\Xi\rm,new}{-}\,\xi_{1\mathbb A}^{\Xi\rm,new}$,\, respectively.
The horizontal gray dashed-lines on the bottom-right panel delimit the 2$\sigma$ range of the data at the bottom of~eq.\,(\ref{ACPexp}).
This translates into the horizontal dashed-lines on the preceding four panels, as explained in the text.
The horizontal purple dotted-lines on each panel delimit the $A_{CP}$ values consistent with collider constraints, as explained in the text.} \label{ACP_vs_Ysd} \vspace{-1em}
\end{figure}

In the top half of figure \ref{ACP_vs_Ysd} we depict the resulting  $A_{CP}^{\Lambda\rm,new}$ in relation to the allowed $|\mathbb Y_{sd,dd,ss}|$ for \,$10^{-4}\le\big|A_{CP}^{\Lambda\rm,new}\big| \le 3.5\times10^{-3}$\, and in the bottom half the corresponding $A_{CP}^{\Xi\rm,new}$  and \,$A_{CP}^{\Lambda\rm,new}{+}A_{CP}^{\Xi\rm,new}$\, versus $|\mathbb Y_{sd}|$.
The aforementioned Re\,$A_2$ bound has turned out to prevent the size of $A_{CP}^{\Lambda\rm,new}$ from surpassing \,$3.5\times10^{-3}$,\, which is well within the 2$\sigma$ range of $A_{CP}^{\Lambda\rm,exp}$ in eq.\,(\ref{ACPexp}).

The top- and bottom-left panels of figure \ref{ACP_vs_Ysd} also exhibit the ranges of the new contributions to the $CP$-violating weak-phase differences in \,$\Lambda\to N\pi$\, and \,$\Xi\to\Lambda\pi$,\, respectively, more exactly \,$-0.027\le\xi_{1\mathbb A}^{\Lambda\rm,new} - \xi_{1\mathbb B}^{\Lambda\rm,new}\le0.027$\, and \,$-0.013\le\xi_{1\mathbb B}^{\Xi\rm,new} - \xi_{1\mathbb A}^{\Xi\rm,new}\le0.013$.\,
Given that the strong-interaction phase shift, \,$\delta_{1\mathbb B} - \delta_{1\mathbb A}$,\, which enters the $A_{CP}$ formula in eq.\,(\ref{acp}), is fairly small and not always precisely known, especially in the $\Xi$ case, as mentioned in section\,\,\ref{hyperons}, it is beneficial to have a direct measurement of \,$\xi_{1\mathbb B}^{} - \xi_{1\mathbb A}^{}$,\, which is less suppressed than $A_{CP}$ due to the absence of the strong phases~\cite{Donoghue:1986hh}.
This has been attempted by BESIII with the $\Xi$ modes~\cite{BESIII:2021ypr,BESIII:2022lsz,BESIII:2023lkg,BESIII:2023drj}, but the results are still compatible with zero and their uncertainties greater than the above corresponding prediction of our NP model.

Our results have also been incorporated into figure \ref{fig:mixcon}, where only the left plot has the permitted points corresponding to \,$10^{-4} \le \big|A_{CP}^{\Lambda\rm,new}\big| < 10^{-3}$ (red) and \,$10^{-3} \le \big|A_{CP}^{\Lambda\rm,new}\big| \le 3.5 \times10^{-3}$ (blue).
We find that the first two conditions in eq.\,(\ref{dm_reqs})
are the main reason for keeping the points from  occupying completely the two lightly-shaded areas in figure\,\,\ref{fig:mixcon}.
Moreover, the Yukawa couplings yielding  \,$\big|A_{CP}^{\Lambda\rm,new}\big|>10^{-3}$\, satisfy \,$|{\rm Im}{\cal\,C}_d/{\rm Im}{\cal\,C}_s-1|\lesssim2\%$,\,
where ${\cal\,C}_{d,s}$ are defined in eq.\,(\ref{C-+u}).

It is evident from the bottom-right panel in figure \ref{ACP_vs_Ysd} that the 2$\sigma$ range of \,$A_{CP}^{\Lambda\rm,exp}{+}A_{CP}^{\Xi\rm,exp}$\, at the bottom of eq.\,(\ref{ACPexp}), delimited by the pair of horizontal gray dashed-lines, already disfavors the bigger values of the prediction, for which \,$A_{CP}^{\Xi\rm,new}\simeq-0.11\,A_{CP}^{\Lambda\rm,new}$.
Taken at face value, this data would suggest for our model the stricter limits
\,$\big|A_{CP}^{\Lambda\rm,new}\big|<1.6\times10^{-3}$\, and \,$\big|A_{CP}^{\Xi\rm,new}\big|<1.7\times10^{-4}$,  which we have indicated in this figure with the horizontal gray dashed-lines on the top panels and bottom-left panel, respectively.

It is worth mentioning that additionally there are constraints on the operators in eq.\,(\ref{L4q}) inferred from collider and other data in global SMEFT analyses~\cite{Bartocci:2024fmm}.
Since ${\cal Q}_{u,\pm}$ are connected to the four-quark operators $Q_{qd}^{(1)}$ and $Q_{qd}^{(8)}$ in the usual SMEFT nomenclature~\cite{Grzadkowski:2010es,Bartocci:2024fmm}, we can employ the limits on their coefficients extracted in ref.\,\cite{Bartocci:2024fmm} to restrict the Yukawa products \,$\mathbb Y_{jk}^{}\mathbb Y_{lo}^*$.\,
We find that the most consequential restriction is \,$|\mathbb Y_{ss}\mathbb Y_{dd}|<0.28$,\, implied by the limit on the $Q_{qd}^{(8)}$ coefficient~\cite{Bartocci:2024fmm}.\footnote{${\cal Q}_{u,\pm}$ are contained in different combinations of \,$\overline{\textit{\texttt q}_j^{}}\textit{\texttt d}_o^{}
\overline{\textit{\texttt d}_l^{}}\textit{\texttt d}_k^{} = -Q_{qd,jklo}^{(1)}/6 - Q_{qd,jklo}^{(8)}$,\, where $j,k,l$, and $o$ are family indices, \,$Q_{qd,jklo}^{(1)} = \overline{\textit{\texttt q}_j^{}}\gamma^{\eta\!}\textit{\texttt q}_k^{}
\overline{\textit{\texttt d}_l^{}}\gamma_\eta^{}\textit{\texttt d}_o^{}$,\, and \,$Q_{qd,jklo}^{(8)} = \overline{\textit{\texttt q}_j^{}} \gamma^\eta\lambda_A^{} \textit{\texttt q}_k^{} \overline{\textit{\texttt d}_l^{}}\gamma_\eta^{}
\lambda_A^{} \textit{\texttt d}_o^{}/4$,\, with $\lambda_A$ being a Gell-Mann matrix.
The coefficients of $Q_{qd}^{(1,8)}$ are $C_{qd}^{(1,8)}/\Lambda^2$ with \,$\Lambda=4$ TeV\, in the notation of ref.\,\,\cite{Bartocci:2024fmm}, which derived the global 95\%-confidence-level limits \,$-13\le C_{qd}^{(1)}\le14$\, and \,$-4.5\le C_{qd}^{(8)}\le7.0$.\bigskip}
This leads to the $A_{CP}^{\rm new}$ ranges confined by the horizontal purple dotted-lines drawn on the plots in figure \ref{ACP_vs_Ysd}.
Although these limits are weaker than those inferred from \,$A_{CP}^{\Lambda\rm,exp}{+}A_{CP}^{\Xi\rm,exp}$\, at the bottom of~eq.\,(\ref{ACPexp}), represented by the dashed lines, the situation may be reversed by future analyses.

\begin{figure}[!b] \bigskip \centering
\includegraphics[width=0.324\linewidth]{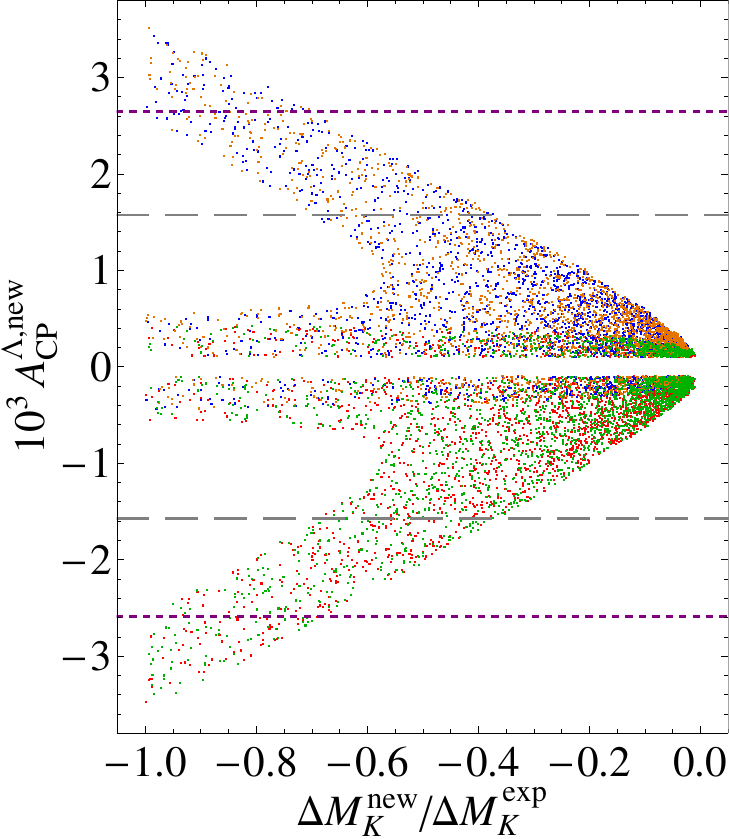}~\includegraphics[width=0.324\linewidth]{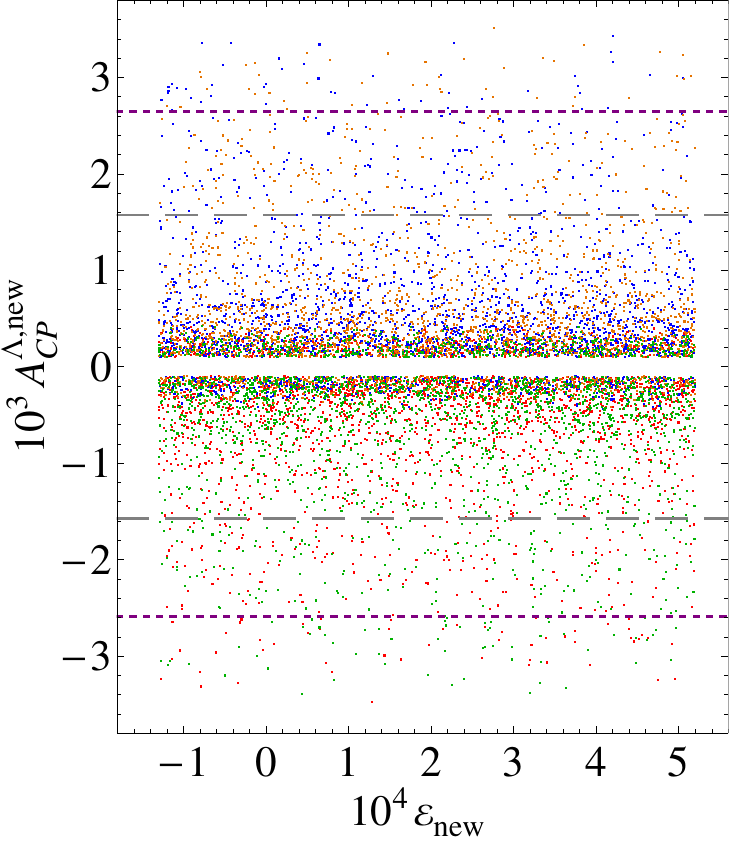}~\includegraphics[width=0.324\linewidth]{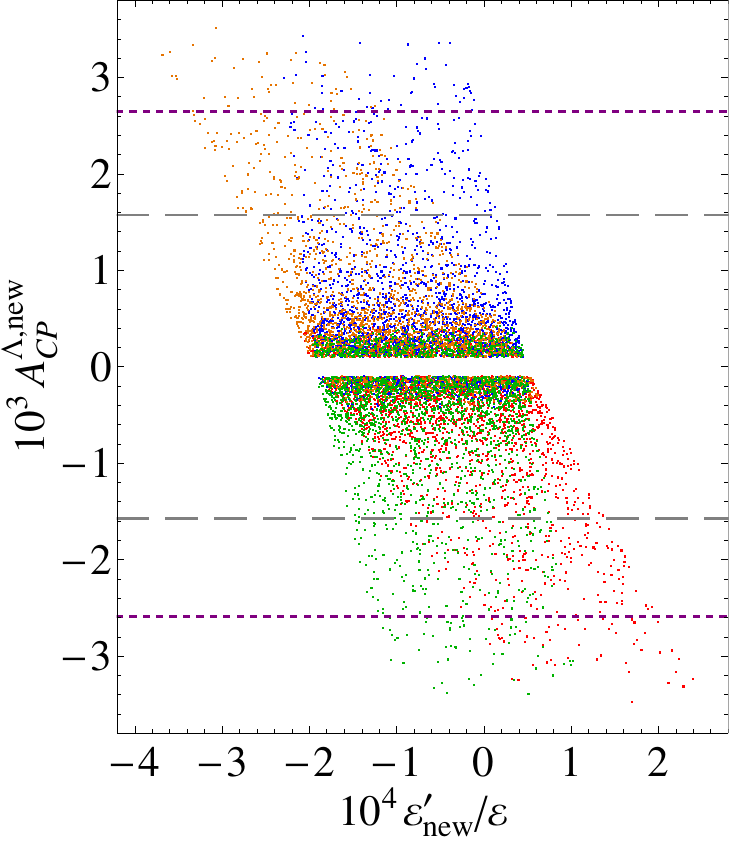}\caption{Distributions of $A^{\Lambda\rm,new}_{CP}$ versus $\Delta M_K^{\rm new}/\Delta M_K^{\rm exp}$ (left), $\varepsilon_{\rm new}$ (middle), and $\varepsilon_{\rm new}'/\varepsilon$ (right), from the Yukawa couplings satisfying the DM and kaon constraints. The horizontal dashed- and dotted-lines are the same as their $A_{CP}^{\Lambda\rm,new}$ counterparts in figure\,\,\ref{ACP_vs_Ysd}.}
\label{fig:ACPL}
\end{figure}

In figure\,\,\ref{fig:ACPL} we display how $A^{\Lambda\rm,new}_{CP}$ correlates with the corresponding new contributions to kaon observables, namely $\Delta M_K^{\rm new}/\Delta M_K^{\rm exp}$, $\varepsilon_{\rm new}$, and $\varepsilon_{\rm new}'/\varepsilon$ on the left, middle, and right plots, respectively.
The left one reveals that $\Delta M_K$ has a role as a limiting factor for the size of $A^{\Lambda\rm,new}_{CP}$.
The middle and right plots indicate that it arises from physics unrestrained by $\varepsilon$ and somewhat less so by $\varepsilon'$.
The horizontal dashed- and dotted-lines in this figure are the same as their counterparts on the  $A^{\Lambda\rm,new}_{CP}$ plots in figure~\ref{ACP_vs_Ysd}.

The proposed Super Tau Charm Facility~\cite{Achasov:2023gey} is anticipated to be capable of probing $A_{CP}^\Lambda$ and $A_{CP}^\Xi$ with  statistical precisions of \,$2.0\times10^{-4}$\, and \,$2.6\times10^{-4}$,\, respectively~\cite{Salone:2022lpt}.
Hence it would potentially be able to test our NP model's predictions for these asymmetries more stringently.
This would expectedly be also feasible~\cite{Schonning:2023mge} in the upcoming PANDA experiment~\cite{PANDA:2020zwv}.
Nearer in the future further efforts may be made by BESIII~\cite{BESIII:2020nme,Zheng:2025tnz}, Belle II~\cite{Belle:2022uod}, and LHCb~\cite{LHCb:2024tnq}.

Finally, it is worth noting that the evaluations of $A_{CP}^{\Lambda\rm,new}$ and $A_{CP}^{\Xi\rm,new}$, which we recall were based on the leading-order chiral realization of ${\cal Q}_{u,\pm}$ given by eqs.\,(\ref{Qu})-(\ref{Qpm}), involve significant uncertainties, possibly up to factors of two.
The situation is similar to the treatment of hyperon nonleptonic decays in chiral perturbation theory where higher-order contributions could be comparable to the lowest-order ones~\cite{Bijnens:1985kj,Jenkins:1991bt,AbdEl-Hady:1999llb}.
This was also seen in the
determination of $A_{CP}$ within the SM~\cite{Tandean:2002vy}.
These uncertainties are implicitly to be added to the $A_{CP}^{\rm new}$ ranges we obtained as well as their derived limits.
It is hoped that lattice-QCD work~\cite{Inoue:2021tdt,Bali:2022qja} in the future could bring about improvements in estimating the asymmetries.

\section{Conclusions\label{concl}}

Motivated by the recent Belle-II finding of the $B^+\to K^+\nu\bar\nu$ rate that is 2$\sigma$ above its standard-model value, we have explored a new-physics scenario, the THDM+D, involving two Higgs doublets possessing tree-level flavor-changing neutral-Higgs interactions and in combination with a real scalar singlet particle, the darkon, acting as a low-mass dark-matter candidate.
Consequently, this model not only can explain the Belle-II anomaly and comply with the requirements from the latest DM relic density and direct search data, but also supplies new sources of $CP$ violation in hyperon and kaon processes.
Imposing restrictions from the DM and kaon sectors, we demonstrated that the resulting $CP$ asymmetries in $\Lambda$- and $\Xi$-hyperon nonleptonic decays could substantially exceed the standard-model expectations and reach levels which are potentially discoverable by ongoing and upcoming experiments.

\section*{Acknowledgements}

J.T. and G.V. thank the Tsung-Dao Lee Institute, Shanghai Jiao Tong University, for kind hospitality and support during different stages of this research.
X.-G.H. was supported by the Fundamental Research Funds for the Central Universities, by the National Natural Science Foundation of the People's Republic of China (No. 12090064, No. 11735010, and No. 11985149), and by MOST 109-2112-M-002-017-MY3.
X.-D.M. was supported by Grant No. NSFC-12305110.
G.V. and X.-G.H. were supported in part by the Australian Government through the Australian Research Council Discovery Project No. DP200101470.

\bigskip\bigskip

\appendix

\section{Hyperon amplitudes\label{ABnew}}

The lowest-order contributions of eq.\,(\ref{Lchi}) to the $\mathbb A$ and $\mathbb B$ constants defined in eq.\,(\ref{m_BB'pi}) for the hyperon decays \,$\Lambda\to p\pi^-,n\pi^0$\, and \,$\Xi^{-,0}\to\Lambda\pi^{-,0}$\, are
\begin{align} \label{Aamplitudes} &
\mathbb A_{\Lambda\to p\pi^-}^{\rm new} \,=\, \frac{ 3\, \tilde{\textsc c}_-^{}\, \hat{\textsc g}_-^{} - \tilde{\textsc c}_+^{}\, \big( 2\hat{\textsc d}_+^{} + 6 \hat{\textsc f}_+^{} + \hat{\textsc g}_+^{} \big) - 2\, \tilde{\textsc c}_u^{}\, \big( \hat{\textsc d}_u^{} + 3\hat{\textsc f}_u^{} - \hat{\textsc g}_u^{} \big) }{2\sqrt3\,f_\pi^{}}
+ \frac{ \sqrt3~ \tilde{\textsc c}_+^{}\, \hat{\textsc h}_+^{}\, (m_\Lambda-m_N)}{f_\pi^3\, \big(m_\pi^2-m_K^2\big)} \,,
\nonumber \\ &
\mathbb A_{\Lambda\to n\pi^0}^{\rm new} \,=\, \frac{ 3\, \tilde{\textsc c}_+^{}\, \hat{\textsc g}_+^{} - \tilde{\textsc c}_-^{}\, \big( 2\hat{\textsc d}_-^{} + 6\hat{\textsc f}_-^{} + \hat{\textsc g}_-^{} \big) }{2\sqrt6\, f_\pi^{}} + \frac{ \sqrt3~ \tilde{\textsc c}_+^{}\, \hat{\textsc h}_+^{}\, (m_\Lambda-m_N)}{\sqrt2\, f_\pi^3\, \big(m_K^2-m_\pi^2\big)} \,,
\nonumber \\ &
\mathbb A_{\Xi^-\to\Lambda\pi^-}^{\rm new} \,=\, \frac{ 3\, \tilde{\textsc c}_-^{}\, \hat{\textsc g}_-^{} - \tilde{\textsc c}_+^{}\, \big( 2\hat{\textsc d}_+^{} - 6 \hat{\textsc f}_+^{} + \hat{\textsc g}_+^{} \big) - 2\, \tilde{\textsc c}_u^{}\, \big(\hat{\textsc d}_u^{}-3\hat{\textsc f}_u^{}-\hat{\textsc g}{}_u'\big) }{2\sqrt3\, f_\pi^{}}
+ \frac{ \sqrt3~ \tilde{\textsc c}_+^{}\, \hat{\textsc h}_+^{}\, (m_\Lambda-m_\Xi)}{f_\pi^3\, \big(m_\pi^2-m_K^2\big)} \,,
\nonumber \\ &
\mathbb A_{\Xi^0\to\Lambda\pi^0}^{\rm new} \,=\, \frac{ 3\, \tilde{\textsc c}_+^{}\, \hat{\textsc g}_+^{} - \tilde{\textsc c}_-^{}\, \big( 2 \hat{\textsc d}_-^{} - 6 \hat{\textsc f}_-^{} + \hat{\textsc g}_-^{}\big) }{2\sqrt6\, f_\pi^{}} + \frac{ \sqrt3~ \tilde{\textsc c}_+^{}\, \hat{\textsc h}_+^{}\, (m_\Lambda-m_\Xi)}{\sqrt2\, f_\pi^3\, \big(m_K^2-m_\pi^2\big)} \,,
\end{align}
\begin{align}
\mathbb B_{\Lambda\to p\pi^-}^{\rm new} & \,=\, \frac{m_\Lambda^{}+m_N^{}}
{\sqrt3\,f_\pi^{}} \!\!\begin{array}[t]{l} \displaystyle \Bigg[ ({\cal D}+{\cal F}) \frac{ 3\, \tilde{\textsc c}_-^{} \hat{\textsc g}_-^{} - \tilde{\textsc c}_+^{} \big( 2\hat{\textsc d}_+^{} + 6 \hat{\textsc f}_+^{} + \hat{\textsc g}_+^{} \big) }{2(m_N-m_\Lambda)}
\\ \displaystyle ~+~ {\cal D}~ \frac{ 2\, \tilde{\textsc c}_+^{} \big(\hat{\textsc d}_+^{}-\hat{\textsc f}_+^{}\big) + \tilde{\textsc c}_u^{} \hat{\textsc g}_u^{} }{m_\Sigma^{}-m_N^{}}
+ \frac{{\cal D}+3{\cal F}}{f_\pi^2} \Bigg( \frac{ \tilde{\textsc c}_+^{} \hat{\textsc h}_+^{} - \tilde{\textsc c}_u^{}\hat{\textsc h}_u^{} }{m_K^2-m_\pi^2} \Bigg) \Bigg] \,, \end{array} ~~~
\nonumber \\
\mathbb B_{\Lambda\to n\pi^0}^{\rm new} & \,=\, \frac{m_\Lambda^{}+m_N^{}}{\sqrt6\,f_\pi^{}} \!\!\begin{array}[t]{l} \displaystyle \Bigg[ ({\cal D}+{\cal F}) \frac{ 3\, \tilde{\textsc c}_-^{} \hat{\textsc g}_-^{} - \tilde{\textsc c}_+^{} \big( 2 \hat{\textsc d}_+^{} + 6 \hat{\textsc f}_+^{} + \hat{\textsc g}_+^{} \big) }{2(m_\Lambda-m_N)}
\\ \displaystyle ~+~ {\cal D}~ \frac{ \tilde{\textsc c}_-^{} \hat{\textsc g}_-^{} + \tilde{\textsc c}_+^{} \big( 2\hat{\textsc d}_+^{} - 2\hat{\textsc f}_+^{} + \hat{\textsc g}_+^{} \big) }{m_N^{}-m_\Sigma^{}} + \frac{ ({\cal D}+3{\cal F})\, \tilde{\textsc c}_-^{} \hat{\textsc h}_-^{} }{f_\pi^2\, \big(m_K^2-m_\pi^2\big)} \Bigg] \,, \end{array}
\nonumber \\
\mathbb B_{\Xi^-\to\Lambda\pi^-}^{\rm new} & \,=\, \frac{m_\Lambda^{}+m_\Xi^{}}{\sqrt3\,f_\pi^{}} \!\!\begin{array}[t]{l} \displaystyle \Bigg[ ({\cal D}-{\cal F}) \frac{ 3\, \tilde{\textsc c}_-^{} \hat{\textsc g}_-^{} - \tilde{\textsc c}_+^{} \big( 2 \hat{\textsc d}_+^{} - 6 \hat{\textsc f}_+^{} + \hat{\textsc g}_+^{} \big) }{2(m_\Xi-m_\Lambda)}
\\ \displaystyle ~+~ {\cal D}~ \frac{ 2\,
\tilde{\textsc c}_+^{} \big( \hat{\textsc d}_+^{} + \hat{\textsc f}_+^{} \big) + \tilde{\textsc c}_u^{} \hat{\textsc g}{}_u' }{m_\Sigma^{}-m_\Xi^{}}
+ \frac{{\cal D}-3{\cal F}}{f_\pi^2} \Bigg( \frac{ \tilde{\textsc c}_+^{} \hat{\textsc h}_+^{} - \tilde{\textsc c}_u^{} \hat{\textsc h}_u^{} }{m_K^2-m_\pi^2} \Bigg) \Bigg] \,, \end{array}
\nonumber \\
\mathbb B_{\Xi^0\to\Lambda\pi^0}^{\rm new} & \,=\, \frac{m_\Lambda^{} + m_\Xi^{}}{\sqrt6\,f_\pi^{}} \!\!\begin{array}[t]{l} \displaystyle \Bigg[ ({\cal D}-{\cal F}) \frac{ 3\, \tilde{\textsc c}_-^{} \hat{\textsc g}_-^{} - \tilde{\textsc c}_+^{} \big( 2\hat{\textsc d}_+^{} - 6\hat{\textsc f}_+^{} + \hat{\textsc g}_+^{} \big) }{2(m_\Lambda-m_\Xi)}
\\ \displaystyle ~+~ {\cal D}~ \frac{ \tilde{\textsc c}_-^{} \hat{\textsc g}_-^{} + \tilde{\textsc c}_+^{} \big( 2\hat{\textsc d}_+^{} + 2\hat{\textsc f}_+^{} + \hat{\textsc g}_+^{} \big) }{m_\Xi^{}-m_\Sigma^{}} + \frac{ ({\cal D}-3{\cal F})\, \tilde{\textsc c}_-^{} \hat{\textsc h}_-^{} }{f_\pi^2\, \big(m_K^2-m_\pi^2\big)} \Bigg] \,, \end{array} \label{Bamplitudes}
\end{align}
where $m_{N,\Sigma,\Xi,\pi,K}$ denote isospin-averaged nucleon, $\Sigma^{+,0,-}$, $\Xi^{0,-}$, $K^{+,0}$, and $\pi^{+,0,-}$ masses, respectively, and \,${\cal D}=0.81\pm0.01$\, and \,${\cal F}=0.47\pm0.01$\, inferred at leading order from the data~\cite{ParticleDataGroup:2024cfk} on semileptonic
octet-baryon decays.
The values of the
$\hat{\textsc d}$'s, $\hat{\textsc f}$'s, $\hat{\textsc g}$'s, and $\hat{\textsc h}$'s are estimated in appendix \ref{dfgh}.

\section{Additional constants\label{dfgh}}

To evaluate the $\hat{\textsc d}$'s, $\hat{\textsc f}$'s, $\hat{\textsc g}$'s, and $\hat{\textsc h}$'s in eqs.\,(\ref{Qu})-(\ref{Qpm}) and appendix \ref{ABnew}, we examine contributions from factorizing the quark operators ${\cal Q}_{u,\pm}$ into products of quark densities.
This entails using the chiral realization of the densities given by~\cite{Tandean:2002vy}
\begin{align} \label{densities}
-\overline{\psi_l^{}}P_L^{}\psi_k^{} & \,\Leftrightarrow~ b_D^{} \big( \xi \big\{ B, \overline B \big\} \xi \big)_{kl} + b_F^{} \big( \xi \big[ B,
 \overline B \big] \xi \big)_{kl} + \tfrac{1}{2} B_0^{} f_\pi^2\, \Sigma_{kl}^{} \,+\, \cdots \,, &
\nonumber \\
-\overline{\psi_l^{}}P_R^{}\psi_k^{} & \,\Leftrightarrow~ b_D^{} \big( \xi^\dagger \big\{ B, \overline B \big\}\xi^\dagger \big)_{kl} + b_F^{} \big( \xi^\dagger \big[ B, \overline B \big] \xi^\dagger \big)_{kl} + \tfrac{1}{2}B_0^{}f_\pi^2\,\Sigma_{kl}^\dagger \,+\,\cdots \,,
\end{align}
where \,$k,l=1,2,3$\, and \,$\psi_{1,2,3}^{} = u,d,s$.\,
Upon applying eq.\,(\ref{densities}) to ${\cal Q}_{u,\pm}$, with \,$b_D^{}=0.226$\, and \,$b_F^{}=-0.811$\, from fitting to the octet baryons
masses,\footnote{We additionally adopt \,$\big(m_u^{},m_d^{},m_s^{}\big)=(2.85,6.20,123)$ MeV\,
at the renormalization scale of 1 GeV.\medskip} \,$B_0^{}=m_{K^0}^2/(m_d+m_s)=1.91$\,GeV,\, and \,$f_\pi^{}=92.07$ MeV\, from ref.\,\cite{ParticleDataGroup:2024cfk}, we arrive at
\begin{align} \label{fact} &
\hat{\textsc d}_u^{} \,=\, \hat{\textsc d}_\pm^{(\prime)} \,=\,  \tfrac{1}{2} b_D^{} B_0^{} f_\pi^2 \,=\, 1.836 \times10^{-3} \rm~GeV^3 \,, &
\nonumber \\ &
\hat{\textsc f}_u^{} \,=\, \hat{\textsc f}_\pm^{(\prime)} \,=\,  \tfrac{1}{2} b_F^{} B_0^{} f_\pi^2 \,=\, -6.576 \times10^{-3} \rm~GeV^3 \,,
\nonumber \\ &
\hat{\textsc h}_{u,\pm}^{} \,=\, \tfrac{1}{4} B_0^2 f_\pi^4 \,=\, 6.571 \times10^{-5} {\rm~GeV^6} \,, ~~~ ~~~~ \hat{\textsc g}{}_u^{(\prime)} = \hat{\textsc g}_\pm^{} \,=\, 0 \,.
\end{align}
Complementarily, we look at nonfactorizable contributions by evaluating the one-hadron matrix elements of ${\cal Q}_{u,\pm}$ in the bag model.\footnote{An introductory treatment of the bag model can be found in \cite{Donoghue:2022wrw}.}
The results are
\begin{align} \label{nonfact} &
\hat{\textsc d}_+^{} =\, \hat{\textsc f}_+^{} =\, 1.484 \times10^{-4} {\rm\,GeV^3} , ~~~ ~
\hat{\textsc g}_u^{} =\,
-3.447 \times10^{-4} {\rm\,GeV^3} , ~~~ ~
\hat{\textsc g}_\pm^{} =\, -2.968 \times10^{-4} {\rm\,GeV^3} ,
\nonumber \\ &
\hat{\textsc h}_u^{} =\, -1.352 \times10^{-6} {\rm~GeV^6} \,, ~~~ ~~~~ \hat{\textsc d}_{u,-}^{} =\, \hat{\textsc f}_{u,-}^{} =\, \hat{\textsc g}{}_u' =\, \hat{\textsc h}_\pm^{} =\, 0 \,,
\end{align}
to be added to their counterparts in eq.\,(\ref{fact}).

\bibliographystyle{utphys.bst}
\bibliography{refs.bib}

\end{document}